%% file: ms-emulapj.tex
\documentclass[12pt,preprint]{emulateapj}
\journalinfo{To appear in the Astrophysical Journal}
\submitted{}

\usepackage{natbib}
\usepackage{lscape}

\newcommand{\fxunits}{\mbox{ergs cm$^{-2}$ s$^{-1}$}}
\newcommand{\lxunits}{\mbox{ergs s$^{-1}$}}

\begin{document}

\title{Spatial Correlation Function of X-ray Selected AGN}

\author{C.R. Mullis\altaffilmark{1},
        J.P. Henry\altaffilmark{2},
	I.M. Gioia\altaffilmark{3},
	H. B\"{o}hringer\altaffilmark{4},\\
	U.G. Briel\altaffilmark{4},
	W. Voges\altaffilmark{4}, and
        J.P. Huchra\altaffilmark{5}}

\altaffiltext{1}{European Southern Observatory, Headquarters,
                 Karl-Schwarzschild-Strasse 2, Garching bei M\"unchen D-85748,
                  Germany, cmullis@eso.org}
\altaffiltext{2}{Institute for Astronomy, University of Hawai`i, 
		 2680 Woodlawn Drive, Honolulu, HI 96822}
\altaffiltext{3}{Istituto di Radioastronomia del CNR, via Gobetti 101, 
                 Bologna, I-40129, Italy}
\altaffiltext{4}{Max-Planck Institut f\"u{r} extraterrestrische Physik,
                 Giessenbachstrasse 1603, Garching, D-85741, Germany}
\altaffiltext{5}{Harvard-Smithsonian Center for Astrophysics, 60 Garden Street, Cambridge, MA 02138}

\shorttitle{CLUSTERING OF X-RAY SELECTED AGN} \shortauthors{MULLIS ET AL.}

\begin{abstract}

We present a detailed description of the first direct measurement of
the spatial correlation function of X-ray selected AGN.  This result
is based on an X-ray flux-limited sample of 219 AGN discovered in the
contiguous 80.7 deg$^{2}$ region of the {\em ROSAT\/} North Ecliptic
Pole (NEP) Survey.  Clustering is detected at the $4\sigma$ level at
comoving scales in the interval $r = 5-60~h^{-1}$ Mpc.  Fitting the
data with a power law of slope $\gamma=1.8$, we find a correlation
length of $r_0 = 7.4^{+1.8}_{-1.9}~h^{-1}$ Mpc ($\Omega_{M}=0.3$,
$\Omega_{\Lambda}=0.7$).  The median redshift of the AGN contributing
to the signal is $z_{\xi}=0.22$. This clustering amplitude implies
that X-ray selected AGN are spatially distributed in a manner similar
to that of optically selected AGN. Furthermore, the {\em ROSAT\/} NEP
determination establishes the local behavior of AGN clustering, a
regime which is poorly sampled in general.  Combined with
high-redshift measures from optical studies, the {\em ROSAT\/} NEP
results argue that the AGN correlation strength essentially does not
evolve with redshift, at least out to $z \sim 2.2$. In the local
Universe, X-ray selected AGN appear to be unbiased relative to
galaxies and the inferred X-ray bias parameter is near unity, $b_{X}
\sim 1$.  Hence X-ray selected AGN closely trace the underlying mass
distribution.  The {\em ROSAT\/} NEP AGN catalog, presented here,
features complete optical identifications and spectroscopic redshifts.
The median redshift, X-ray flux, and X-ray luminosity are $z=0.41$,
$f_{\rm X}=1.1 \times 10^{-13}$ \fxunits, and $L_{\rm X}=9.2 \times
10^{43}~h_{70}^{-2}$ \lxunits~(0.5--2.0 keV), respectively.
Unobscured, type 1 AGN are the dominant constituents (90\%) of this
soft X-ray selected sample of AGN.

\end{abstract}


\section{Introduction} 
\label{Introduction} 

Active galactic nuclei (AGN) are presumably the very luminous
manifestations of accretion onto supermassive black holes.
Shining brightly across the electromagnetic spectrum and easily
detectable to very high redshift, AGN are accessible tracers of galaxy
formation and evolution as well as large-scale structure.  The spatial
distribution of AGN reflects the distribution of matter fluctuations
modulated by the complex, non-linear astrophysics of black hole
formation.  Thus measurements of AGN clustering and its evolution
provide important tests for models of AGN formation in an adopted
cosmological reference frame \citep{Hartwick1990}.

The vast majority of work on this topic has focused on optical surveys
for quasi-stellar objects (QSOs).  Following the first attempt to
measure AGN clustering by \citet{Osmer1981}, the first significant
detection came from the work of \citet{Shaver1984}.  In the ensuing
years sample sizes have increased as has the resulting precision
\citep{Shanks1987,Iovino1988,Andreani1992,Mo1993,Shanks1994,
Croom1996,LaFranca1998}.  There is broad agreement in the conclusions
of these studies.  Clustering is detected at the $\sim$3--4$\sigma$
level.  The spatial distribution is characterized using the
conventional approach of the two-point correlation function which
tracks the excess probability over random to find two objects
separated by a given distance.  This function is found to have a
power-law shape with a slope of $\gamma \approx 1.8$.  The clustering
scale length, where the correlation function is unity, is $r_0 \sim
6~h^{-1}$ Mpc measured at a mean redshift of $\bar{z} \sim 1.4$.  This
amplitude is comparable to that found for luminous, local galaxies,
and larger than that of dwarf or low surface brightness galaxies
(e.g., the results from the Sloan Digital Sky Survey:
\citealt{Zehavi2002} and the 2dF Galaxy Redshift Survey:
\citealt{Hawkins2003}).  This is consistent with the idea that
moderately powerful AGN are found in moderately luminous galaxies, and
so share their clustering properties.

One question left unresolved by the aforementioned papers is the
nature of the clustering evolution, the redshift-dependent behavior of
the scale length. \citet{Croom2001} have used the 2dF QSO Redshift
Survey (2QZ) to definitively redress this issue, at least in the high
redshift interval, and have found essentially no change in the
clustering amplitude between $z\sim 0.5$ and \mbox{$z\sim$ 2.2}
\citep[see also][]{Croom2003}.  The recent measure by
\citet{Grazian2004} of the local AGN population lends further support
to this largely flat behavior.

At X-ray wavelengths there are far fewer results on the clustering of
AGN chiefly due to a historic lack of sufficiently large X-ray
samples, particularly those with comprehensive optical follow-up.  The
{\em ROSAT} all-sky survey (1990--1991), the first with an imaging
X-ray detector, and the subsequent program of pointed observations
enabled the first significant steps forward.  Despite the currently
limited development, this field is anticipated to experience rapid
growth with the advent of much deeper surveys from {\em Chandra}, {\em
XMM-Newton} and future X-ray survey missions.

The use of X-rays in the selection and characterization of AGN
is a natural choice.  \mbox{X-ray} emission appears to be a universal
feature of AGN \citep{Elvis1978}, and essentially all optically
selected AGN are X-ray luminous \citep{Avni1986}.  In fact, X-ray
emission is likely the least biased selection technique, particularly
at hard energies above 2 keV \citep{Mushotzky2004}.

There have been several determinations of the angular correlation of
X-ray AGN \citep{Vikhlinin1995,Akylas2000,Giacconi2001,Basilakos2004}.
However, translating these results to constraints on the
three-dimensional clustering strength requires a number of assumptions
that results in only weak constraints on $r_{0}$.  Actually measuring
the spatial correlation function is more involved since optical
identifications and redshift measures are required.  Early attempts by
\citet{Boyle1993} and \cite{Carrera1998} using data from {\em Einstein} and
{\em ROSAT\/}, respectively, did not yield significant clustering
signals.

The first direct measure of the spatial correlation function of X-ray
selected AGN was achieved in the context of the {\em ROSAT\/} North
Ecliptic Pole (NEP) Survey.  This result was initially reported by
\citet{Mullis2001b}.  In the current paper we provide a detailed
description of this clustering analysis.  Note that only very recently
have additional measures of this kind become available as reported for
the {\em Chandra} deep fields by \citet{Gilli2004}.

We describe the {\em ROSAT\/} NEP Survey in \S\,\ref{data} focusing
on the basic properties of the X-ray flux-limited AGN sample and the
survey selection function.  In \S\,\ref{analysis} the formalism of the
clustering analysis is laid out, including a description of the spatial
correlation estimator and the Monte Carlo simulations of the
survey volume.  We present the spatial correlation function of \mbox{X-ray}
selected AGN in \S\,\ref{results}.  These results are compared to
similar studies at both X-ray and optical wavelengths with emphasis on
the evolution of clustering strength with redshift in
\S\,\ref{discussion}.  We close with a summary of the key results in
\S\,\ref{conclusions}.

Unless otherwise stated, we use a cosmological model with the
parameters, $H_{0} = 70~h_{70}$ km s$^{-1}$ Mpc$^{-1}$,
$\Omega_{M}=0.3$, and $\Omega_{\Lambda} = 0.7$, and refer to this as
$\Lambda$CDM.  However, for the correlation length we use $H_{0} =
100$ h$^{-1}$ km s$^{-1}$ Mpc$^{-1}$ for ready comparison to previous
results.  Unabsorbed X-ray fluxes and restframe X-ray luminosities are
quoted in the \mbox{ 0.5--2.0 keV} energy band, unless otherwise
indicated.  Measurement errors are given at the 68\% confidence
interval ($1\sigma$).

\section{The ROSAT NEP AGN Sample}
\label{data}

\begin{figure*}
\epsscale{1}
\plotone{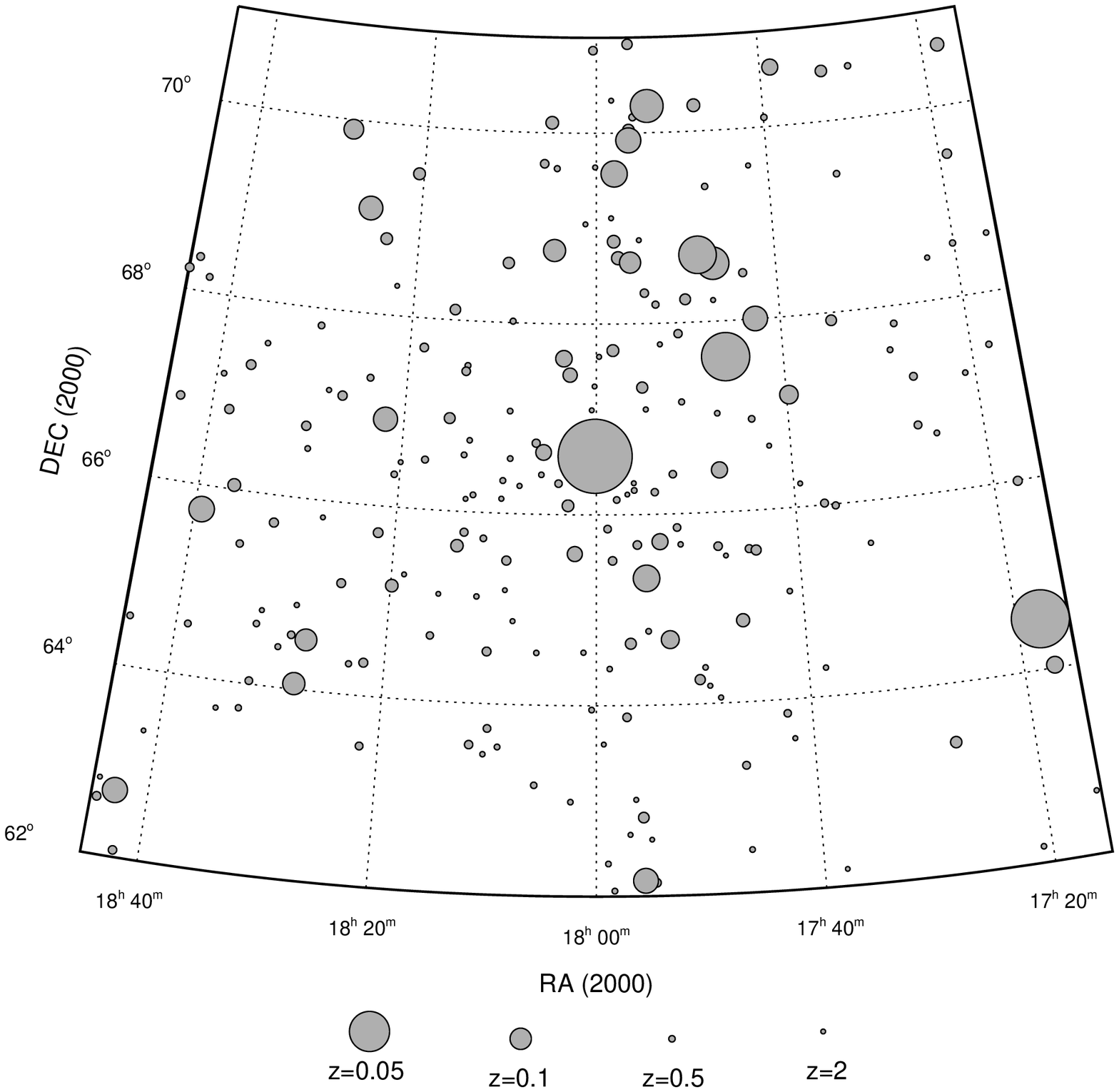}
 
\caption{Distribution on the sky of the 219 AGN in the {\em ROSAT\/}
NEP Survey (Aitoff projection).  To give some perception of the
three-dimensional arrangement of the sample, the symbol size reflects
the projected size of a constant physical radius of 500 $h^{-1}$ kpc
at the redshift of the AGN.  As the X-ray emitting region of AGN is
sub-parsec, this scale is greatly exaggerated for visualization
purposes.  An animated ``fly-through'' of the {\em ROSAT\/} NEP Survey
volume is available at {\tt http://www.ifa.hawaii.edu/\~{
}mullis/nep3d.html}.\label{fig:agnskyplot}}
 
\end{figure*}

In the {\em ROSAT\/} All-Sky Survey (RASS), an 80.7 deg$^{2}$ region
around the NEP (\mbox{$\alpha_{2000}=18^{\rm h}00^{\rm m}$},
\mbox{$\delta_{2000}=+66^{\circ}33$\arcmin}) constitutes one of the
deepest observations of the X-ray sky ever achieved with such a large,
contiguous solid angle \citep{Mullis2001b,Henry2001,Voges2001}.  Here
445 unique X-ray sources are detected with fluxes measured at greater
than $4\sigma$ significance in the 0.5--2.0 keV energy band.  The
median and maximum exposure times are approximately 5 and 38 ks,
respectively. We have identified the physical nature of 443 (99.6\%)
of the {\em ROSAT\/} NEP X-ray sources through a comprehensive program
of imaging and spectroscopy \citep{Gioia2003}. AGN are the dominant
constituents comprising nearly half (49.4\%) of identified sources in
the {\em ROSAT\/} NEP Survey catalog \mbox{(Figure \ref{fig:agnskyplot})}.

The {\em ROSAT\/} NEP AGN are particularly well suited for a
clustering analysis.  The AGN are drawn from a contiguous, wide-angle
region in the sky sampled to a relatively deep X-ray flux.  The survey
selection function is well determined and is only a function of X-ray
flux.  We have identified all but two of the 445 X-ray sources in the
survey region.  Thus the AGN sample is essentially complete and
requires no complicated assumptions to correct for
incompleteness. Furthermore, we have spectroscopically measured
redshifts for the entire sample.

\subsection{Basic Properties}

We present the full sample of 219 {\em ROSAT\/} NEP AGN in
{\mbox{Table \ref{tab:agn}}.  This includes several notable revisions
relative to previous versions of the catalog
\citep{Mullis2001b,Gioia2003}.  First, the AGN fluxes and
luminosities previously reported were over-estimated by approximately
20\% on average due to an error in the conversion of X-ray count rate
to flux.\footnote{A power-law photon index of $\Gamma$=1 was
mistakenly used \mbox{instead} of $\Gamma$=2 due to a typographical error in
the analysis software.  \mbox{X-ray} fluxes and luminosities for
galaxy clusters and stars in the referenced works are not affected by
this problem.} Second, the sample has grown by one due to the
re-classification of an X-ray source (RX\,J1824.7+6509; see footnote
in {\mbox{Table \ref{tab:agn}}).  And finally, we have adopted the
presently-favored ``concordance'' cosmology in computing X-ray
luminosities.  The revised and updated catalog with corrected
properties is presented in this paper and should be the reference
point in any future work with the {\em ROSAT\/} NEP AGN sample.  Below
we explain the columns of {\mbox{Table \ref{tab:agn}}.\\

Columns (1) and (2): the object name and internal identification
number.  Sources are listed in order of increasing right ascension.

Columns (3) and (4): the right ascension and declination of the X-ray
centroid, respectively.

Columns (5) and (6): the right ascension and declination of the
optical counterpart, respectively.

Column(7): Column density of Galactic hydrogen from
\citet{Elvis1994} with supplements from \citet{Stark1992}.

Columns (8) and (9): Count rate and count rate error in the
\mbox{0.1--2.4 keV} energy band measured within a circular aperture of
5\arcmin~radius.  The quoted error is the 1$\sigma$ uncertainty based on a
maximum-likelihood analysis.

Column (10): Unabsorbed flux in the \mbox{0.5--2.0 keV} \mbox{energy} band
derived from the count rate assuming the source has a power-law
spectrum (photon index $\Gamma=2$), with absorption fixed at the
Galactic value for the source position.  This total flux includes a
correction factor of 1.0369 which accounts for the small fraction of
point source flux falling outside the photometry aperture.

Column (11): Rest-frame, K-corrected luminosity in the \mbox{0.5--2.0
keV} energy band.  Note the K-correction for a power-law spectrum is
$(1+z)^{\Gamma-2}$ and is thus unity for $\Gamma=2$ (cosmology:
$H_{0} = 70~h_{70}$ km s$^{-1}$ Mpc$^{-1}$, $\Omega_{M}=0.3$, and
$\Omega_{\Lambda} = 0.7$).

Column (12): Spectroscopic redshift (typical uncertainty $<0.001$).

Column (13): AGN classification based on the equivalent width of emission
lines and the broadness of permitted lines (1 = type 1, 2 = type 2).\\

\begin{figure*}
\epsscale{1}
\plotone{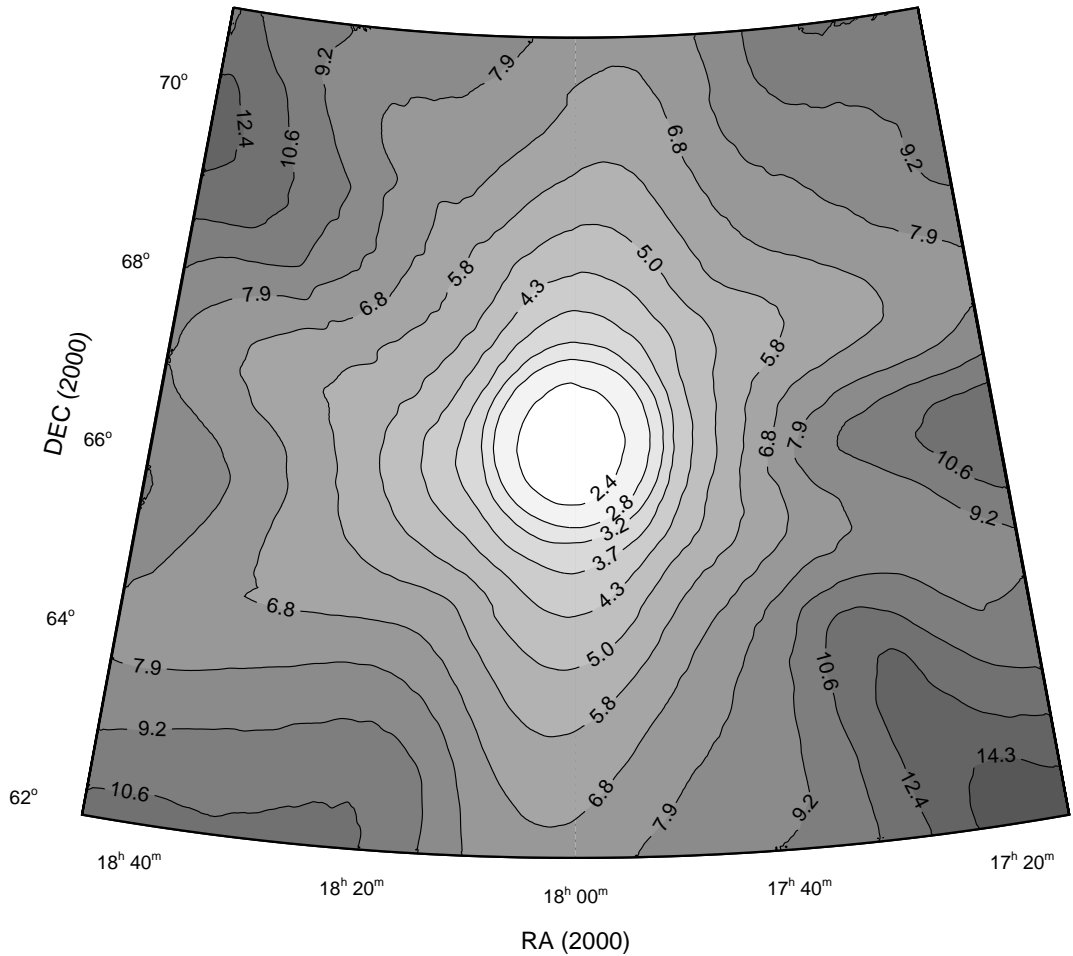}
 
\caption{Map of the X-ray flux limits for the detection of AGN in the
{\em ROSAT\/} NEP Survey (Aitoff projection).  The
logarithmically-spaced contours indicate the minimum total flux that
an object must have to meet the $4\sigma$ detection requirement of the
survey.  Contours are labeled in units of $10^{-14}$ \fxunits~ in the
0.5--2.0\,keV energy band.  The enhanced sensitivity toward the
center of the survey region is a result of the {\em ROSAT\/} scan
pattern that converges at the NEP.\label{fig:fluxmap}}
 
\end{figure*}

We have secured spectroscopic redshifts for the entire {\em ROSAT\/}
NEP AGN sample; the full range is $z=0.026-3.89$ and
the median is $z=0.41$. X-ray luminosities range over
10$^{41}-10^{46}$ \lxunits~ with a median of $L_{\rm X}=9.2
\times 10^{43}~h_{70}^{-2}$ \lxunits. Since {\em ROSAT\/} has very limited
sensitivity to hard X-rays above 2 keV, the bulk of the NEP AGN (90.4\%, 198
objects) are type I AGN (QSOs and Seyfert 1 galaxies) based
on the equivalent width and the broadness of their permitted emission
lines \mbox{($W_{\lambda}\ge 5$ \AA, FWHM $\ge 2000$ km s$^{-1}$)}.
The remaining 21 NEP AGN (9.6\%) are categorized as type II (Seyfert 2
and star-forming galaxies). 

\subsection{Sky Coverage and logN($>$S)-logS Distribution}
\label{logNlogS}

The distribution on the sky of the 219 {\em ROSAT \/} NEP AGN is shown
in \mbox{Figure \ref{fig:agnskyplot}}.  The average AGN source density
is \mbox{2.7 deg$^{-2}$} and increases toward the NEP at the center of
the survey region as a result of the RASS scan pattern.  During the
survey phase of the mission, {\em ROSAT\/} scanned the sky in great
circles which overlapped at the ecliptic poles.  The resulting peak in
integrated exposure time at the NEP is reflected in the sensitivity
map shown in \mbox{Figure \ref{fig:fluxmap}}.  This map of limiting
X-ray flux is derived from RASS exposure and background maps (see
Chapter 3 of \citealt{Mullis2001b} and J.\@ P.\@ Henry et al. 2004, in
preparation). Here the contours, in units of $10^{-14}$ \fxunits,
indicate the minimum flux that an AGN must have to produce at least a
$4\sigma$ detection and thus meet the selection criterion of the {\em
ROSAT\/} NEP survey.  The sensitivity map is integrated to obtain the
final selection function shown in \mbox{Figure \ref{fig:skycov}}.  The
effective sky coverage is 81 deg$^{2}$ at bright fluxes, starts to
decrease below $1.3 \times 10^{-13}$ \fxunits, and remains significant
down to $\sim 2.3 \times 10^{-14}$ \fxunits~ where the coverage is 1
deg$^{2}$.

In subsequent modeling of the AGN population we will make use of the
integral number counts or \mbox{$\log N(>S)-\log S$} distribution of
the {\em ROSAT\/} NEP AGN sample.  We calculate the number of objects
per square degree observed above a flux $S$ by summing up the
contribution of each source, weighted by the area in which the source
could have been detected, via the equation,

\begin{equation}
N(>S) = \sum_{S_i > S} \frac{1}{\Omega(S_i)},
\label{lognlogs}
\end{equation}

\noindent where $S_i$ is the flux of the $i$th source and $\Omega(S_i)$ is the
sky coverage at flux $S_i$ from \mbox{Figure \ref{fig:skycov}}. A
maximum likelihood fit of a power law \citep*{Murdoch1973} to the
differential source counts over the flux range
\mbox{(2--1000)~$\times~10^{-14}$} \fxunits gives the relation

\begin{equation}
N(>S) = (1.5 \pm 0.1) \left(\frac{S}{10^{-13}}\right)^{-1.30 \pm 0.08}~\rm{deg^{-2}}.
\end{equation}

\noindent The normalization is set by the total number of AGN observed
above the minimum survey flux.  The measured \mbox{$\log N(>S)-\log
S$} and the power-law fit are plotted in \mbox{Figure
\ref{fig:lognlogs}}.  These results are consistent with previous
determinations in this flux regime. For example, \citet{Hasinger1993}
found a slope of \mbox{$1.72 \pm 0.27$} at fluxes \mbox{$\ga
2.7~\times~10^{-14}$ \fxunits,} and \citet{Mason2000} report a slope
of \mbox{$1.6 \pm 0.3$} at fluxes above \mbox{$3~\times~10^{-14}$
\fxunits.}

\section{Clustering Analysis}
\label{analysis}

One of the simplest and most popular techniques for characterizing the
clustering of objects is the two-point spatial correlation function
\citep{Peebles1980}.  This statistic is defined in terms of the joint
probability ($dP$) of simultaneously finding an object in a volume $dV_{1}$
and another object in a volume $dV_{2}$ separated by a distance $r$,

\begin{equation}
dP ~=~ n^2\, [1 + \xi(r)]\ dV_{1} dV_{2},
\label{cfdef}
\end{equation}

\noindent where $n$ is the average number density of objects and
$\xi(r)$ is the two-point spatial correlation function.  In a uniform
random Poisson point process, objects are distributed in a completely
random pattern and the probabilities of finding objects in $dV_{1}$
and $dV_{2}$ are independent such that $\xi(r) = 0$.  If objects are
more clustered than average $\xi(r) > 0$, whereas, if objects are more
dispersed than average $\xi(r) < 0$.  Hence, the two-point correlation
is the excess probability over random to find two objects separated by
distance $r$.  

%

\begin{figure}
\epsscale{1.2}
\plotone{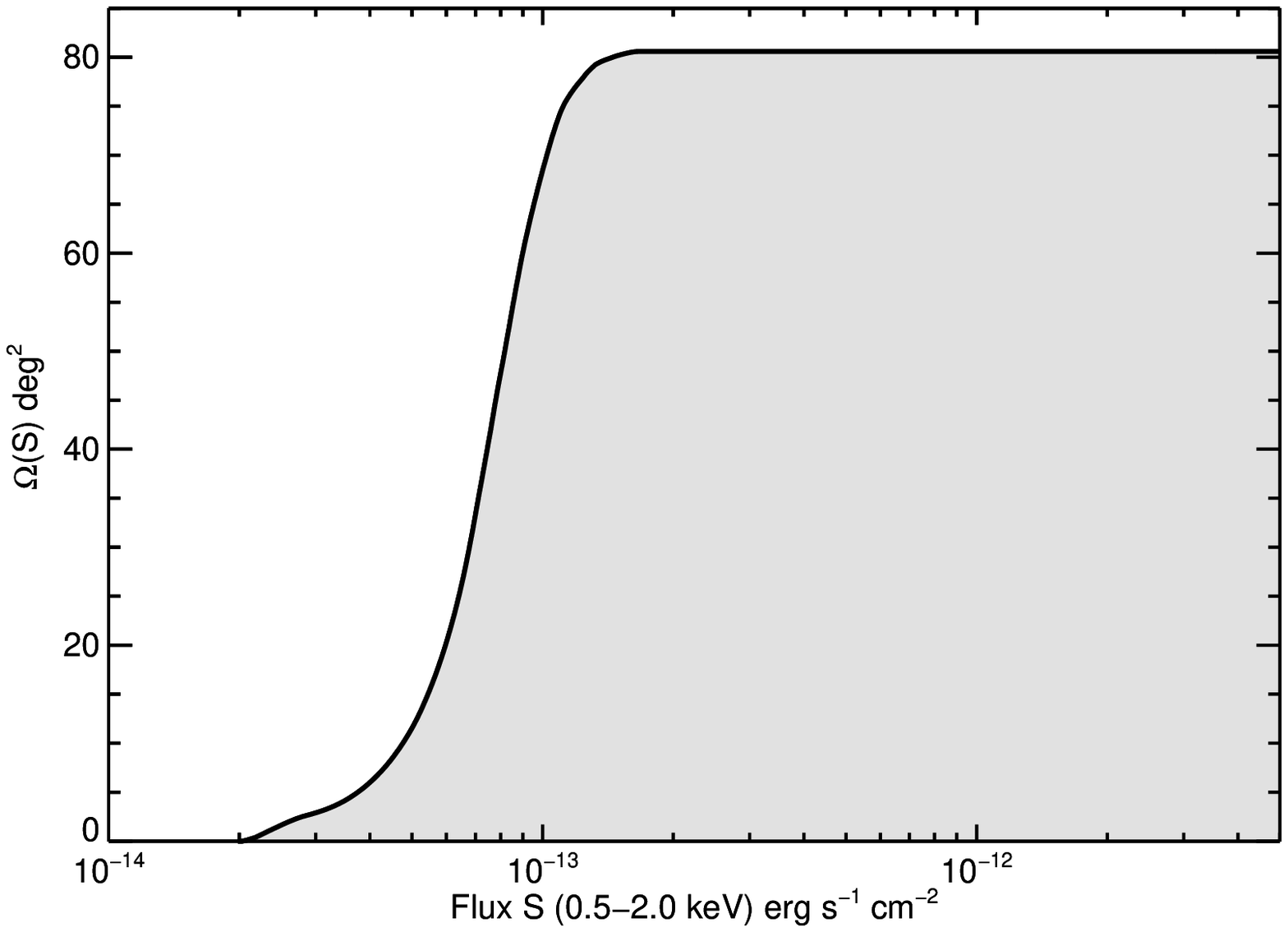}
 
\caption{Sky coverage for AGN in the {\em ROSAT\/} NEP Survey.  This
is the total area of the sky surveyed as a function of total flux.\label{fig:skycov}}
 

\epsscale{1.2}
\plotone{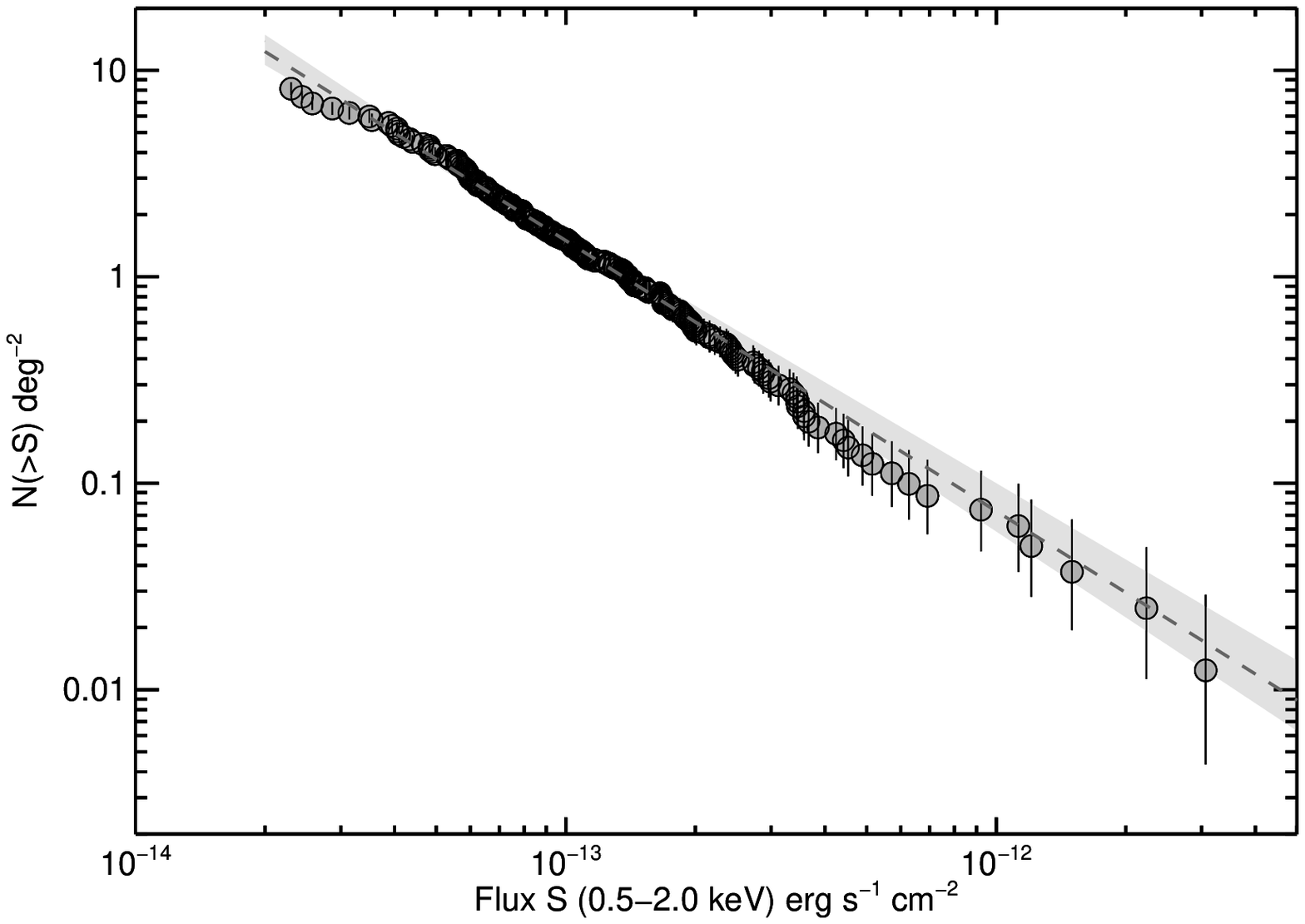}
 
\caption{Integral number counts per square degree of AGN in the {\em
ROSAT\/} NEP Survey.  The dashed line is a power-law fit to the data
derived from a maximum likelihood treatment of the differential number
counts.  The shaded region is the $1\sigma$ error region for this
fit. \label{fig:lognlogs}}
 
\end{figure}

The observed correlation function for a variety of extragalactic
objects (e.g., galaxies, AGN, and galaxy clusters) is well fit by a
power law of the form

\begin{equation}
\xi(r) ~ = ~ \left( \frac{r}{r_{0}} \right)^{-\gamma}
\label{cf}
\end{equation}

\noindent where the slope is typically \mbox{$\gamma$~$\approx$~1.8}
while the correlation length $r_{0}$ depends on the type of object.

\subsection{Correlation Estimator}

The correlation function is extracted from spatial data via a pairwise
analysis.  In essence, the number of object pairs of a given spatial
separation in the data are compared to the number of object pairs of
the same separation from a random catalog.  The random catalog is
created by homogeneously populating the survey volume in a manner that is
consistent with the selection function and boundary conditions of the
survey.  Hence, an enhancement of the number of data-data pairs
relative to the corresponding random-random pairs is indicative of
structure in the data.  

Though the correlation function can be computed using the natural form,
\mbox{$\xi_{n}(r)= (DD/RR)-1$}, \citet{Landy1993} have demonstrated that the
variance is minimized using the estimator,

\begin{equation}
\xi(r)~=~ \frac{N_{\rm r}(N_{\rm r}-1)}{N_{\rm d}(N_{\rm d}-1)} 
          \frac{DD(r)}{RR(r)} - 
          \frac{(N_{\rm r}-1)}{N_{\rm d}} \frac{DR(r)}{RR(r)} + 1.
\label{cfest}
\end{equation}

\noindent Here $DD(r)$ is the number of data-data pairs in the NEP AGN
sample with separations of $r \pm \Delta r/2$ in redshift space.
Similarly, $RR(r)$ is the number of random-random pairs and $DR(r)$ is
the number of data-random cross-pairs, each with separations of $r \pm
\Delta r/2$.  The number of objects in the data sample is $N_{\rm d}$,
and the number of objects in the random sample is $N_{\rm r}$.  A
number of different correlation estimators are commonly used
\citep[e.g.,][]{Davis1983,Hewett1982,Hamilton1993}, but a detailed
comparison by \citet*{Kerscher2000} shows that the \mbox{Landy \&
Szalay} estimator is preferred in terms of minimal bias and variance,
and gives the most reliable results on all spatial scales.


We calculate co-moving separations $r$ between objects in redshift
space using the standard relations
\citep[e.g.,][]{Weinberg1972,Peebles1993,Hogg1999}.  The transverse
co-moving distance between the observer and an object at redshift $z$
is

\begin{eqnarray}
D_{\rm M}(z)~=~ & & \frac{\rm c}{H_0\sqrt{|\Omega_{k}|}}~ \times \nonumber \\
& & {\cal{S}}\left( \sqrt{|\Omega_{k}|} \int_0^z dz'
[\Omega_M(1+z')^3~+~\right. \nonumber\\ 
& & \hspace{8mm} \left. \Omega_{k}(1+z')^2~+~\Omega_{\Lambda}]^{-1/2}\right)
\label{rdist}
\end{eqnarray} 
where $\Omega_{{M}}$ is the matter density parameter, and
$\Omega_{\Lambda}$ is the cosmological constant.  The curvature of
space is characterized by \mbox{$\Omega_{{k}} = 1 -
\Omega_{M} - \Omega_{\Lambda}$}.  The function ${\cal{S}}$ is
defined according to:
\[ {\cal{S}}(x) \equiv \left\{ \begin{array}
   {r@{\quad:\quad}ll}
   \sinh{x} & \Omega_{k} > 0 & {\rm open~universe}\\ 
   x        & \Omega_{k} = 0 & {\rm flat~universe}\\
   \sin{x}  & \Omega_{{k}} < 0 & {\rm closed~universe} 
 \end{array} \right. \]

\noindent Note that the transverse co-moving distance ($D_{\rm M}$) is
related to the luminosity distance ($D_{\rm L}$) and the angular
diameter distance ($D_{\rm A}$) via the equations \mbox{$D_{\rm
L}=(1+z) D_{\rm M}$} and \mbox{$D_{\rm A}=D_{\rm M}/(1+z)$}.

 Consider two objects separated on the sky by an angle
$\theta$ with transverse co-moving radial distances of $D_{\rm M1}$
and $D_{\rm M2}$.  The co-moving separation of the second object as measured from the first object is

\begin{equation}
r~=~\sqrt{d^2 D_{\rm M1}^2 + D_{\rm M2}^2 - 2\, d\, D_{\rm M1}\, D_{\rm M2} \cos{\theta}},
\label{cosine}
\end{equation}

\noindent where

\begin{eqnarray}
\lefteqn{d~=~\sqrt{1 + \Omega_{{k}} \left( \frac{H_{0}\,D_{\rm M1}}{{\rm c}}\right)^2}~+}\nonumber\\
& & ~\frac{D_{\rm M1}\cos{\theta}}{D_{\rm M2}}  \left(1 - \sqrt{1 + \Omega_{{k}} \left( \frac{H_{0}\,D_{\rm M2}}{{\rm c}}\right)^2}\ \right)\  
\end{eqnarray}
\citep[e.g.,][]{Osmer1981,Matarrese1997}.
\noindent In a flat universe $\Omega_{{k}} = 0$, $d=1$ and Equation
\ref{cosine} reduces to the cosine rule for Euclidean space.  Note
that Equation \ref{cosine} is not symmetric for the positions of the
two objects.  However, this is only important in the case in which $d$
is not close to unity.

\subsection{Construction of Random Samples}
\label{random_construction} 

Generating random samples for the survey volume is critical for
estimating the correlation function.  The objective is to construct
simulated datasets with constituents that are consistent with the
physical distribution of the source population and are selected in the
same manner as the real data.  In the following Monte Carlo procedure
we create an X-ray flux-limited sample of simulated AGN possessing the
same flux and redshift distributions as the actual {\em ROSAT\/} NEP
AGN.  Of course the important difference between the simulated and
real AGN is that the parent population of the former is assumed to be
randomly distributed on the sky and in redshift space.

To properly model the non-uniform sensitivity pattern of the {\em
ROSAT\/} NEP survey region \mbox{(Figure \ref{fig:fluxmap})}, the
simulated AGN must have the same flux distribution as the real AGN.
We have demonstrated in \S\,\ref{logNlogS} that the observed AGN flux
distribution follows a power law of the form \mbox{$N(>S)=K
S^{-\alpha}$} where $\alpha \approx 1.3$ \mbox{(Figure
\ref{fig:lognlogs})}.  Thus the differential probability distribution
of fluxes scales like \mbox{$S^{-(\alpha + 1)}$}.  In practice the
required set of random numbers with a power-law distribution is
obtained from a set of random uniform numbers using a transformation
method \citep[e.g.,][]{Bevington1992}. If $p$ is a random number
uniformly distributed between 0 and 1, then a random sampling in X-ray
flux $S$ above a limiting flux $S_{\rm lim}$ is distributed like

\begin{equation}
S~=~S_{\rm lim}\ (1 - p)^{-1/\alpha}.
\label{S}
\end{equation}

\noindent The limiting flux in the {\em ROSAT\/} NEP Survey for AGN is
\mbox{$S_{\rm lim}=2.0 \times 10^{-14}$ erg s$^{-1}$ cm$^{-2}$}. Small
variations to the value of $\alpha$ do not change the outcome of the
clustering analysis.

Redshifts for the simulated AGN are drawn from a probability density
function based on the observed distribution of the {\em ROSAT\/} NEP
AGN \mbox{(Figure \ref{fig:zdistr}}).  Here the redshift distribution
has been smoothed with a Gaussian of width $\sigma_z=0.1$. This
smoothing scale is selected such that the resulting distribution has
the overall shape of the data but is not strongly affected by discrete
large-scale structure or Poisson fluctuations.  Our results are
insensitive to reasonable variations in this smoothing length.

We randomly place simulated AGN within the {\em ROSAT\/} NEP Survey
boundaries (\mbox{$17^{h}15^{m} < \alpha_{2000} < 18^{h}45^{m}$},
\mbox{$62^{\circ} < \delta_{2000} < 71^{\circ}$}).  For each object we
assign an X-ray flux according to Equation \ref{S} and a redshift
drawn from the smoothed distribution.  Then we test to see if its flux
is above the local flux limit for its particular position in the
survey region as shown in \mbox{Figure \ref{fig:fluxmap}}.  Objects
above the threshold are retained, while those below are rejected.  We
proceed to randomly populate the survey volume in this manner until we
have constructed a random catalog of 100,000 simulated AGN; i.e.,
$\sim450$ times the size of the {\em ROSAT\/} NEP AGN sample.

\begin{figure}
\epsscale{1.2}
\plotone{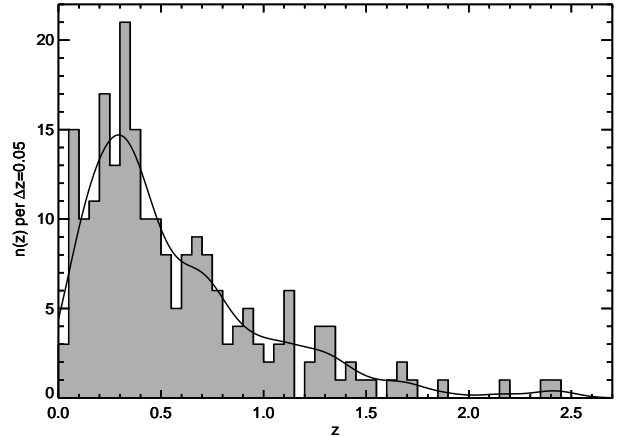}
 
\caption{Redshift distribution of the {\em ROSAT\/} NEP AGN.  The
black line is the distribution smoothed with a Gaussian of width
$\sigma_z=0.1$.  The median redshift of the sample is
$z=0.41$. To preserve the readability of this plot, the most
distant AGN at \mbox{$z=3.889$} is not shown.\label{fig:zdistr}}
 
\end{figure}

\section{Results}
\label{results}

\begin{figure*}
\epsscale{1}
\plotone{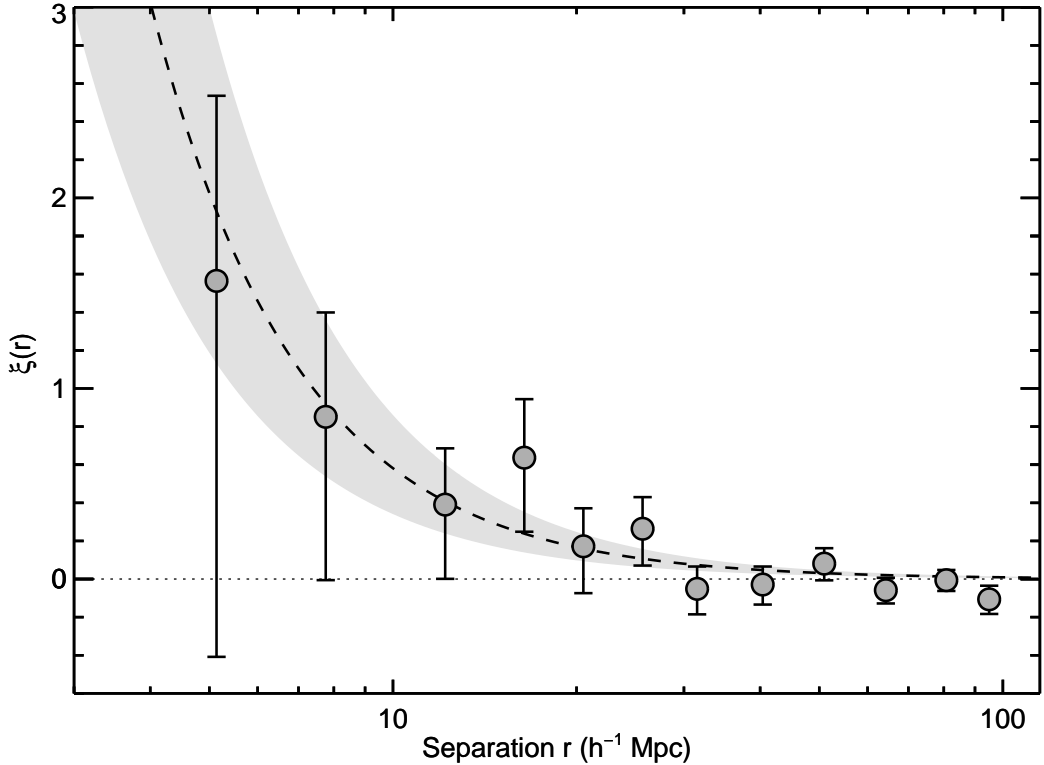}
 
\caption{Spatial correlation function of the {\em ROSAT\/} NEP
AGN. The error bars on the data points are $1\sigma$ Poisson errors.
The dashed line is the maximum-likelihood fit
($r_0=7.4^{+1.8}_{-1.9}\,h^{-1}$ Mpc, $\gamma=1.8$ fixed) to the range
from \mbox{5 to 60\,$h^{-1}$ Mpc} ($\Omega_{M}=0.3$ and
$\Omega_{\Lambda} = 0.7$).  The shaded region demonstrates the $\pm
1\sigma$ errors on the correlation strength.\label{fig:agncf}}

 \end{figure*}

We present the spatial correlation function of the {\em ROSAT\/} NEP
AGN in Figure \ref{fig:agncf}.  Here the data and simulated objects
have been binned up in terms of their pair separations (e.g., $DD(r)$,
$RR(r)$, $DR(r)$) and evaluated using the adopted correlation
estimator \mbox{(Equation \ref{cfest})}.  The uncertainties associated
with these data points are typically estimated using Poisson statistics
of the form,

\begin{equation}
\delta \xi(r) ~=~ \frac{1 + \xi(r)}{\sqrt{DD(r)}}
\label{basicerror}
\end{equation}

\noindent where $DD(r)$ is the number of data-data pairs in the
interval.  However, the Poisson distribution is not well approximated
by a Gaussian in situations where the number of counts in each bin is
small ($N \la 20$) and thus the $\sqrt{N}$ errors in the denominator
of \mbox{Equation \ref{basicerror}} can underestimate the 68\%
confidence level.  To avoid this we use the formulas of
\citet{Gehrels1986} to estimate the Poisson confidence intervals for
one-sided 84\% upper and lower bounds which correspond to $\pm1\sigma$
in Gaussian statistics.  

A positive clustering signal is readily apparent in the spatial
correlation function of {\em ROSAT\/} NEP AGN and is significant at
the $4\sigma$ level based on the excess AGN pairs relative to random
at $r \la 60~h^{-1}$ Mpc.  Thus it is desirable to estimate the
correlation strength and slope using the canonical power-law fit.
This could be done using the coarsely binned data in \mbox{Figure
\ref{fig:agncf}} and minimizing the ${\chi}^2$ statistic.  However,
this approach is subject to uncertainties because the determination
can be sensitive to the size and distribution of the selected bins.
The maximum-likelihood method is an alternative means for determining
a power-law fit to the correlation function which has the advantage of
making maximal use of the data and is free from arbitrary binning
(e.g., \citealt{Croft1997,Popowski1998}; \citealt*{Borgani1999};
\citealt{Moscardini2000,Collins2000}).  The co-moving separation $r$
is parsed into very small intervals such that there is either 0 or 1
object pair in any given interval.  In this limit Poisson
probabilities are appropriate.  The probability $P$ of observing
$\nu_{i}$ object pairs where $\mu_{i}$ pairs are expected is given by

\begin{equation}
P_{\mu_{i}}(\nu_{i}) ~=~\frac {e^{-\mu_{i}}{\mu_{i}}^{\nu_{i}}}{\nu_{i}!}.
\label{poission_p}
\end{equation}

\noindent In the sparse sampling limit, the probabilities associated
with the bins are independent of each other such that a likelihood
function ${\cal{L}}$ can be defined in terms of the joint
probabilities,

\begin{equation}
{\cal{L}}~=~\prod_{i} \frac{{\rm e}^{-\mu_{i}} \mu_{i}^{-\nu_{i}}}{\nu_{i}!}.
\label{mleq}
\end{equation}

\noindent This leads to the useful expression
\begin{equation}
\ln{{\cal{L}}}~=~\sum_{i} \left( -\mu_{i} +\nu_{i} \ln{\mu_{i}} 
                  - \ln{\nu_{i}!} \right).
\label{sumform}
\end{equation}

\noindent Here the summation is over all the intervals $r_{i}$ within
the range of co-moving separations for which the power-law fit is to
be determined.  The number of observed object pairs $\nu_{i}$ are the
data-data pairs measured from the {\em ROSAT\/} NEP AGN, whereas the
expected numbers of pairs $\mu_{i}$ are calculated using the
\mbox{Landy \& Szalay} estimator in Equation \ref{cfest}.  Notice that
$\mu_{i}$, which is $DD(r_{i})$ in Equation \ref{cfest}, is a function
of the power-law parameters $r_{0}$ and $\gamma$ and the pairwise data
from the observed and random samples, specifically $DR(r_{i})$ and
$RR(r_{i})$.

The best-fitting values of the power-law fit to the correlation
function are determined by minimizing the expression

\begin{equation}
S=-2\ ln{\cal{L}}
\end{equation}
with the confidence levels defined to be 
\begin{equation}
\Delta S ~=~  S(r_{0},\gamma) - S(r_{\rm 0,best},\gamma_{\rm best}).
\end{equation}

\noindent Since $S$ is distributed like $\chi^2$, the $1\sigma$,
$2\sigma$, and $3\sigma$ (68.3\%, 95.4\%, and 99.7\%) confidence
intervals for a two parameter fit are $\Delta S = 2.30$, 6.17, and
11.8, respectively \citep{Avni1976,Cash1976,Cash1979}. If we fix the
slope, the confidence intervals for the one-parameter fit are $\Delta
S = 1.0$, 4.0, and 9.0.

We show the results of the maximum-likelihood analysis of the {\em
ROSAT\/} NEP AGN clustering data in \mbox{Figure \ref{fig:ml}}.  The
power-law fit over the range from \mbox{5 to 60\,$h^{-1}$ Mpc}
indicates best-fit values of \mbox{$r_0=7.5^{+2.7}_{-4.2}\,h^{-1}$
Mpc} and \mbox{$\gamma=1.85^{+1.90}_{-0.80}$} where the errors are
$1\sigma$ for two interesting parameters.  If we set the slope at
$\gamma = 1.8$ (the value typically found for normal galaxies and
optically selected AGN), the correlation length is \mbox{$r_{0} = 7.4
^{+1.8}_{-1.9}~h^{-1}$ Mpc}.  We overplot this best-fit power-law on
the pairwise data in \mbox{Figure \ref{fig:agncf}}.  The lower limit
on the fitting range is set by the smallest AGN-AGN separation in the
NEP data; the upper limit matches the general break in the power-law
shape of the observed function.  Our results do not change
significantly for reasonable variations in the fitting range.  For
example, varying the upper limit between \mbox{$40~h^{-1}$ Mpc} and
\mbox{$80 ~h^{-1}$ Mpc} results in the best-fit $r_0$ for $\gamma=1.8$
ranging between \mbox{$7.3~h^{-1}$ Mpc} and \mbox{$7.6~h^{-1}$ Mpc}
(i.e., $\Delta r_0 \la 0.2 \sigma$).  Note that the effective redshift
of the NEP clustering signal is $z_{\xi}=0.22$ given by the median
redshift of the AGN in pairs with separations in the range
\mbox{5--60\,$h^{-1}$ Mpc}.  This characteristic redshift is smaller
than the sample median (z=0.41) since the density of detected objects,
and thus the number of close pairs, decreases with redshift in this
flux-limited survey.

\begin{figure}
\epsscale{1.2}
\plotone{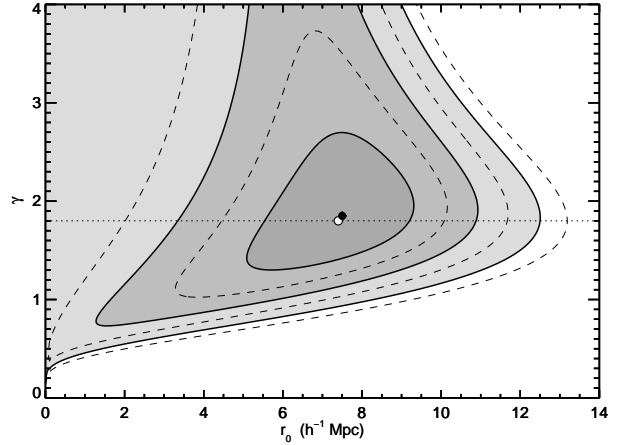}
 
\caption{Probability contours for the power-law normalization ($r_{0}$)
and slope ($\gamma$) from a maximum-likelihood analysis of the {\em
ROSAT\/} NEP AGN ($\Omega_{M}=0.3$ and $\Omega_{\Lambda} = 0.7$).  The
solid contours with shading are the $1\sigma$, $2\sigma$, and
$3\sigma$ (68.3\%, 95.4\%, and 99.7\%) confidence intervals for a one
parameter fit.  The dashed contours are the corresponding intervals
for a two parameter fit.  The filled circle marks the best-fit for two
free parameters, \mbox{$r_0=7.5^{+2.7}_{-4.2}\,h^{-1}$ Mpc} and
\mbox{$\gamma=1.85^{+1.90}_{-0.80}$}; the open circle marks
the best-fit for one free parameter with the slope fixed at
$\gamma=1.8$, \mbox{$r_{0} = 7.4 ^{+1.8}_{-1.9}\,h^{-1}$ Mpc}. The
dotted horizontal line indicates the canonical power-law slope for
the correlation function ($\gamma \approx 1.8$).\label{fig:ml}}

\end{figure}

Repeating this analysis assuming an Einstein-de Sitter cosmology
($\Omega_{M} = 1$, $\Omega_{\Lambda} = 0$), we find best-fit values of
\mbox{$r_0=7.2^{+2.3}_{-4.0}\,h^{-1}$ Mpc} and
\mbox{$\gamma=1.83^{+1.66}_{-1.00}$}, and \mbox{$r_{0} = 7.1 \pm
1.6~h^{-1}$ Mpc} with $\gamma = 1.8$. 
This modest decrease in the correlation
length is expected since the effect of a significant
$\Omega_{\Lambda}$ is to increase object-object separations $r$ and
thus the resulting correlation length $r_0$. 
These results are in excellent agreement with the preliminary analysis
of the NEP data first reported in \citet{Mullis2001b}.

There is a substantial supercluster of galaxies at \mbox{$z=0.087$} in
the {\em ROSAT\/} NEP Survey region \citep{Mullis2001a}.  This is
manifested as $5\sigma$ spikes in the redshift distributions of galaxy
clusters and {\em IRAS\/} galaxies, but only a $2\sigma$ fluctuation
in AGN density.  There are 12 AGN from our sample in the redshift
regime of the NEP supercluster ($0.07 < z < 0.1$).  These objects
comprise 5\% of the total sample and 12\% of the AGN at $z < 0.4$.  We
have examined the potential impact of the superstructure on the
clustering signal by restricting the analysis to the 200 AGN at $z >
0.1$.  This result is entirely consistent with the results from the
analysis of the full sample (\mbox{$\Delta r_{0} \sim 0.6\sigma$,}
\mbox{$\Delta \gamma \sim 0.2\sigma$)}.  Thus the NEP supercluster has
no significant effect on our results.

A final concern lies in the potential consequences of source confusion
as a result of the angular resolution of the {\em ROSAT\/} PSPC
detector.  The FWHM of the RASS PSF is $\sim$35\arcsec, and thus we
expect sources separated by less than $\sim$30\arcsec~ to be
unresolved.  However, the confusion-limited regime is never approached
due to the relative sparsity of AGN at the sensitivity limits of the
{\em ROSAT\/} NEP Survey.  For example, the smallest distance between
two NEP AGN is 3.8\arcmin.  Furthermore, at the typical redshift of
AGN contributing to the correlation detection ($z_{\xi}=0.22$),
30\arcsec~ corresponds to a co-moving separation of $56~h^{-1}$
kpc which is much smaller than the best-fit correlation length.  Thus
confusion effects should not bias the correlation analysis.

\section{Discussion}
\label{discussion}

In this section we discuss the significance of the {\em ROSAT\/} NEP
results for the clustering of AGN, make comparisons to similar
measures, and discuss the implications of our results for the
evolution of AGN clustering.

\subsection{First Direct Measure of the Spatial Correlation Function 
of X-ray Selected AGN}
\label{direct}

Our results from the analysis of {\em ROSAT\/} NEP data
represent the first direct and significant measure of the spatial
correlation function of X-ray selected AGN.  This outcome is due to
the advantageous combination of relatively deep sensitivity over a
wide, contiguous survey region which results in sufficient numbers of
AGN-AGN pairs at small scales to yield a significant signal.  For
example, the maximal signal-to-noise ratio in the pairwise data is
observed at co-moving separations of less than \mbox{30 $h^{-1}$
Mpc}. Here we find 123 AGN-AGN pairs where only 78 pairs are expected
if the data had a uniform spatial distribution.  Thus AGN clustering
is detected at greater than $4\sigma$ level. 

The correlation length derived from the {\em ROSAT\/} NEP AGN sample
is consistent with the value $r_{0}\sim 6$ $h^{-1}$ Mpc associated
with normal galaxies \citep[e.g.,][]{Hawkins2003} and optically
selected AGN \citep[e.g.,][]{Croom2001}.  This suggests that X-ray
luminous AGN are spatially clustered in a manner similar to that of
the aforementioned objects.  In fact it lends additional evidence to
the idea that \mbox{X-ray} selected and optically selected AGN are
drawn from the same population.  Our observations imply that AGN are
not biased relative to galaxies based on the similarity of clustering
strengths.  Note that the factor of two density enhancement observed
in {\em both} AGN and galaxies within the NEP supercluster provides
further support to this view \citep[see Figure 1 in][]{Mullis2001a}.
As observation and theory indicate the bias of local galaxies is close
to unity \citep[e.g.,][and references
therein]{Verde2002,Weinberg2004}, the inferred bias parameter for
X-ray selected AGN is also near unity, $b_{X} \sim 1$.  Previous
measurements of the X-ray bias, some of which included all
extragalactic sources, have varied widely \citep[see ][and references
therein]{Barcons2001}; however our NEP measurement is in good
agreement with recent results based on the hard X-ray background
\citep{Boughn2004}.

It is pertinent to review the two efforts to measure the spatial
correlation of X-ray selected AGN prior to that of the NEP.  The fact
that there are such few studies reflects the difficulty of constructing
sufficiently large and appropriately distributed samples to make this
measure.  The first clustering analysis of this type was performed by
\citet{Boyle1993} using 183 AGN ($z < 0.2$) taken from the {\em Einstein\/}
Extended Medium Sensitivity Survey (EMSS; \citealt{Stocke1991}).
Though this sample is similar in size to the NEP, it is spread over
a nearly ten times larger solid angle \mbox{(770 deg$^{2}$)}.
Moreover, their survey region is not contiguous, but rather the combination
of hundreds of pointings distributed across the sky.
\citet{Boyle1993} found the significance of the clustering signal
was only $0.8\sigma$ and concluded that AGN clustering at small scales
is very weak, if present at all.

The second work to pursue the spatial correlation function is that of
\citet{Carrera1998} who examined two separate samples of X-ray
selected AGN.  The first consisted of 107 objects ($0 < z < 3.5$,
$\bar{z}$ = 1.43) from the {\em ROSAT\/} Deep Survey (DRS;
\citealt{Shanks1991}).  The second was composed of 128 AGN ($0 < z <
3.5$, $\bar{z}$ = 0.84) from the {\em ROSAT\/} International X-ray
Optical Survey (RIXOS; \citealt{Mason2000}).  Like the EMSS, the solid
angles surveyed by the DRS and RIXOS (1.4 deg$^2$ and 20 deg$^{2}$,
respectively) are the summation of many individual pointed
observations. \citet{Carrera1998} found no indication of
clustering in the DRS sample, but detected clustering at $r < 40$--80
h$^{-1}$ Mpc in the RIXOS sample at about $1.7\sigma$ significance.
Given the limitations of the data, the authors did not present an
explicit correlation function.  However, by combining the DRS and
RIXOS AGN, they derived a correlation length in the range of $1.5 <
r_{0} < 5.5~h^{-1}$ Mpc dependent upon the assumed model of clustering
evolution.  These results for an Einstein-de Sitter cosmological
model would increase by approximately 30\% in the concordance model.

Aside from our {\em ROSAT\/} NEP results, the only other significant
measure of this kind to date is from the recent analysis of the {\em
Chandra} deep fields. \citet{Gilli2004} have measured the projected
correlation function ($w(r_{p}$) where $r_p$ is the co-moving
separation perpendicular to the line of sight) using the 2Msec {\em
Chandra} Deep Field North (CDFN) and 1Msec {\em Chandra} Deep Field
South (CDFS).  This work is based on 160 AGN from the CDFN (mean
redshift, ${z} \sim 0.96$) and 97 AGN from the CDFS ($z \sim 0.84$).
Though these samples are smaller than that of the NEP, they are
concentrated in much smaller solid angles, $\sim 0.1$ deg$^{2}$ each.
Thus \citeauthor{Gilli2004} are able to extract high signal-noise
results by probing the correlation function at very small scales
\mbox{($\sim 0.2-10~h^{-1}$ Mpc)} where the signal is comparatively
stronger.  Converting the projected correlation to the three-dimension
scale length assuming a power-law shape, they find for the CDFN
\mbox{$r_{0}=5.5 \pm 0.6~h^{-1}$ Mpc}, \mbox{$\gamma=1.50 \pm 0.12$};
for the CDFS \mbox{$r_{0}=10.3 \pm 1.7~h^{-1}$ Mpc},
\mbox{$\gamma=1.33 \pm 0.14$}.  If the slope is fixed at their
preferred value of $\gamma=1.4$, they find for the CDFN
\mbox{$r_0=5.1^{+0.4}_{-0.5}~h^{-1}$ Mpc} and for the CDFS
\mbox{$r_0=10.4 \pm 0.8~h^{-1}$ Mpc}.  There is a large variance
between the two fields.  However the strong enhancement in the
correlation strength of the CDFS is attributed to two large redshift
spikes at $z \sim 0.7$.

It is enlightening to compare the {\em ROSAT\/} NEP and CDFN/S spatial
correlation functions.  The clustering strength of the NEP,
\mbox{$r_{0} \approx 7.4~h^{-1}$ Mpc}, lies intermediate in the
$r_{0}=5-10 ~h^{-1}$ Mpc range of the {\em Chandra} deep fields. The
{\em ROSAT} AGN are a soft X-ray \mbox{(0.1--2.4\,keV)} selected
sample compared to the {\em Chandra} samples assembled from detections
at both soft to hard energies \mbox{(0.5--10\, keV)}.  Consequently
the {\em ROSAT\/} NEP sample is dominated by type 1 AGN, whereas those
of CDFN/S are composed of nearly equal mixes of \mbox{type 1} and
\mbox{type 2} AGN.  However, \citet{Gilli2004} have measured the
correlation strength separately for hard versus soft detected AGN as
well as type 1 versus type 2 AGN and find no significant differences.
In relative terms, the NEP AGN are luminous and nearby while the
CDFN/S AGN are faint and distant.  The NEP characterizes the local ($z
\sim 0.2$) clustering of AGN versus the distant ($z \sim 0.9$)
measures of the CDFN/S, a point we will elaborate on in
\S\,\ref{sect:evol}.  The median luminosity of the {\em ROSAT\/} NEP
AGN is \mbox{$L_{\rm X} = 9.2 \times 10^{43}~h_{70}^{-2}$
\lxunits} whereas the mean of the CDFN/S is \mbox{$L_{\rm X} = 6.0
\times 10^{42}~h_{70}^{-2}$ \lxunits}~(both in the 0.5--2.0 keV energy band).

\subsection{Comparison to the Angular Clustering of X-ray Sources}

Thus far we have restricted our discussion to the direct measure of
the spatial correlation of AGN using X-ray flux-limited samples.
There are very few investigations of this kind due to the
observational challenges of securing X-ray data of sufficient depth
and/or breadth, and obtaining optical follow-up observations for
object identifications and spectroscopic redshifts.  If we allow for
additional assumptions and uncertainties, the {\em angular}
correlation function of X-ray sources can be incorporated.

\begin{figure*}
\epsscale{1}
\plotone{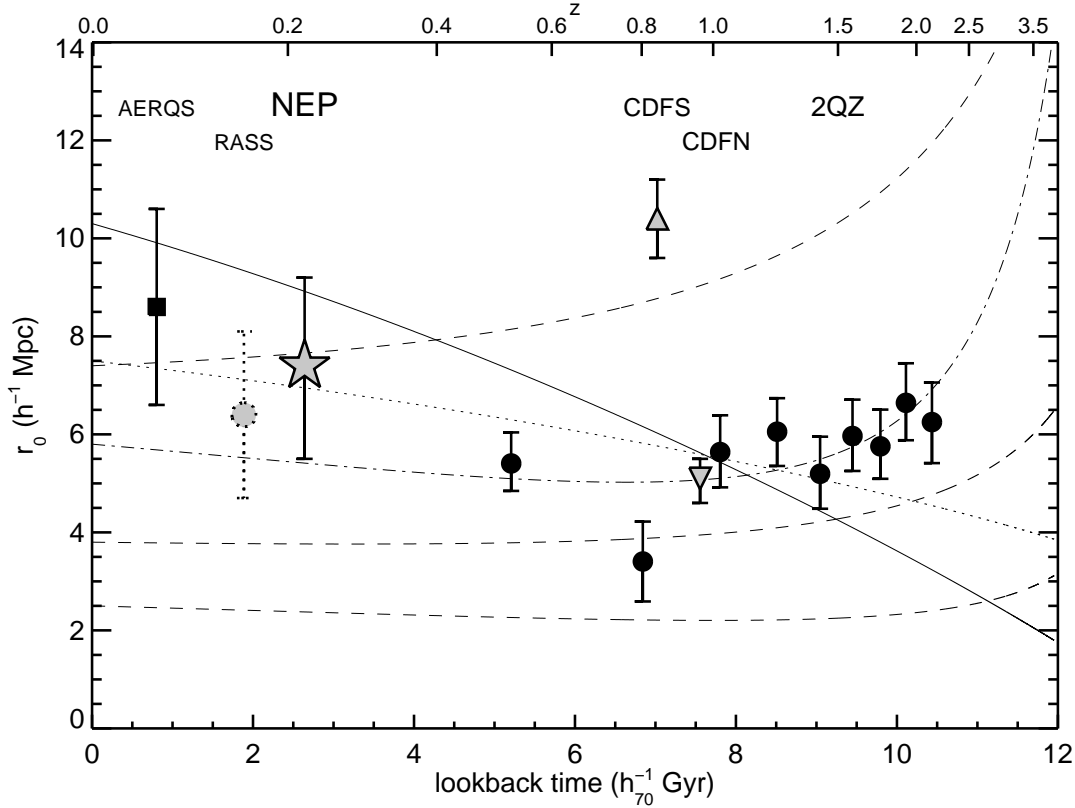}
 
\caption{The scale length of AGN clustering as a function of lookback
time (lower axis) and redshift (upper axis).  X-ray results are shaded
symbols; optical results are solid symbols.  Symbol key and references
for the data are: AERQS \citep[filled square;][]{Grazian2004}, RASS
\citep[shaded circle;][]{Akylas2000}, NEP \citep[shaded star;][and
this paper]{Mullis2001a}, CDFN/S \citep[shaded
triangles;][]{Gilli2004}, and 2QZ \citep[filled
circles;][]{Croom2003}. All of these are from direct measures of the
spatial correlation function except for the RASS.  Over-plotted are
theoretical predictions based on different biasing models described by
\citet[][and references therein]{Croom2001}: linear theory (solid
line), long-lived AGN (dotted line), merger model for different values
of the minimum halo mass (dashed line, 10$^{12}$\,M$_\sun$,
10$^{13}$\,M$_\sun$, 10$^{14}$\,M$_\sun$ bottom to top).  The
dot-dashed line is the best-fitting empirical bias model for the 2QZ
data.  The uncertainty intervals of the measurements are based
on one-parameter fits.  The NEP and RASS use a fixed slope of
$\gamma=1.8$; AERQS and 2QZ assume $\gamma=1.56$.  CDFN/S use their
preferred value of $\gamma=1.4$. All of the measurements and models
are for a $\Lambda$CDM cosmological model.
\label{fig:r0evol}}
 
\end{figure*}

The angular clustering of X-ray sources has been detected several
times
\citep[e.g.,][]{Vikhlinin1995,Akylas2000,Giacconi2001,Basilakos2004}.
The bulk of this signal is presumably generated by AGN since they are
the dominant class of X-ray emitters at high Galactic latitudes.
However, in the absence of optical follow-up, the exact AGN fraction
is uncertain.  The completely identified {\em ROSAT\/} NEP Survey
provides a reference point at soft energies (0.5--2.0\,keV) and at
moderately faint fluxes (a few times 10$^{-14}$ \fxunits) where
$\sim$50\% of the X-ray sources are AGN.  A second important caveat
concerning these angular studies is that both the redshift
distribution of the sources and the redshift dependence of the
clustering evolution must be assumed to extract the three-dimensional
correlation length $r_0$.

\citet{Vikhlinin1995} reported the first detection of the angular
clustering of discrete X-ray sources using a set of deep {\em ROSAT\/}
observations.  They concluded that the correlation strength is
consistent with that of optically selected AGN assuming such objects
constitute a large fraction ($\ga$50\%) of their X-ray sources.
\citet{Akylas2000} measured the angular clustering of X-ray sources
from the RASS Bright Source Catalog, taking precautions to exclude
stellar and extended sources as much as possible.  Assuming a typical
redshift of $z\sim 0.1-0.2$, comoving clustering evolution,
and a power-law slope of $\gamma \approx 1.8$, they derived a correlation
length of $r_0 \approx 6.4 \pm 1.7~h^{-1}$ Mpc; thus similar to the
results at high redshift from optical surveys.  We have converted
their EdS result to $\Lambda$CDM assuming a median redshift of
$z=0.15$.

Recently \citet{Basilakos2004} analyzed the clustering of {\em
XMM-Newton} hard X-ray sources (2--8\,keV) in a \mbox{2 deg$^{2}$}
region.  Exploring a variety of potential luminosity functions (and
thus redshift distributions) and clustering characteristics, they found
a rather high correlation length spanning the range $r_0 \sim
9-19~h^{-1}$ Mpc for $\gamma=1.8$.  \citeauthor{Basilakos2004} argue
that hard-band sources are more strongly correlated than soft-band
sources.  This claim is supported by the counts-in-cell analysis from
deep {\em Chandra} data by \citet{Yang2003}, but is at odds with the
results of \citet{Gilli2004} for the CDFN/S.

\subsection{Evolution of AGN Clustering}
\label{sect:evol}

To round out this discussion, we incorporate the latest results on the
clustering of {\em optically selected} AGN, and use the ensemble of
X-ray and optical data to constrain the evolution of AGN clustering.

As outlined in \S\,\ref{Introduction}, there is a significant body of
work in the literature characterizing the clustering of optically
selected AGN.  The current state of the art is encapsulated in the
work of \citet{Croom2001,Croom2003} with the 2QZ survey.  Analyzing a
sample of over 20,000 AGN, they find the two-point spatial correlation
has a power-law shape over the range $1-60~h^{-1}$ Mpc.  Their
best-fit parameters are $r_0=5.76^{+0.17}_{-0.27}~h^{-1}$ Mpc and
$\gamma=1.64^{+0.06}_{-0.03}$ ($\Lambda$CDM) measured at a mean
redshift of redshift $\bar{z} \simeq 1.5$.  Moreover, the 2QZ sample
is sufficiently large to map out the variation in clustering strength
at high redshifts.  \citet{Croom2003} find no evolution in the
clustering amplitude from $z \sim 0.5$ to $z \sim 2.2$.

Ironic in the era of high-redshift studies, the clustering properties
of {\em local} AGN has only recently been measured directly
\citep{Mullis2001b,Grazian2004}.  Our {\em ROSAT\/} NEP analysis has
an effective redshift of $z_{\xi} \sim 0.2$.  \citet{Grazian2004} have
determined a similar measure for optically selected AGN using data
from the Asiago-ESO/RASS QSO Survey (AERQS).  They find an amplitude
of $r_0 = 8.6 \pm 2.0~h^{-1}$ Mpc ($\Lambda$CDM) at an effective
redshift of $z_{\xi}=0.06$.  The AERQS and NEP results are in
excellent agreement.

A composite view of the clustering evolution of AGN is shown in
\mbox{Figure \ref{fig:r0evol}}.  Here we plot the scale length of the
spatial correlation as a function of redshift and lookback time; X-ray
results are shaded symbols, optical results are solid
symbols.  Note that all of these are direct measures of
the spatial correlation function except for the RASS data point
\citep{Akylas2000} which is a transformation of the angular signal.  
Any conclusions to be drawn from these data must be taken with due
regard for several caveats that will be discussed shortly.  However,
for now we take the data points at face value.

To the zeroth order no evolution is observed in the clustering
strength of AGN out to redshift $z \sim 2.2$.  A straight line at $r_0
= 6~h^{-1}$ Mpc is a good fit to the data with only two points
deviating by greater than $3\sigma$.  The CDFS is an obvious outlier,
but, as noted before this high value is likely due to cosmic variance.
Looking with more detail, a weighted mean of the local measures (NEP,
RASS, and AERQS) indicates \mbox{$r_{0} = 7.3 \pm 1.0~ h^{-1}$
Mpc}; whereas the high-redshift results (2QZ and CDFN/S) have a
weighted mean of \mbox{$r_{0} = 5.80 \pm 0.21~h^{-1}$ Mpc}.  Thus
there is an $\sim$25\% increase in clustering at low redshift ($z \sim
0.2$) relative to high redshift ($z \approx 1-2$).  However, this very
mild redshift evolution is significant at only the 1.4$\sigma$ level.

The conclusion of zero to weak negative evolution is generally
applicable to either the X-ray or optically selected AGN on their own,
though with much less confidence for the former.  Our {\em ROSAT\/}
NEP results and the {\em Chandra} deep fields provide complementary
redshift coverage.  However, the trend based on strictly these X-ray
AGN is still open ended due to the high variance of CDFN/S results.
Forthcoming analysis at high redshift from {\em Chandra} and {\em
XMM-Newton} will hopefully remove this uncertainty.  We anticipate
future findings will favor correlation lengths of $r_0 \approx
6~h^{-1}$ Mpc on average.

Our interpretation of the clustering behavior shown in \mbox{Figure
\ref{fig:r0evol}} must be tempered by several potentially important
issues, including: 1) X-ray versus optical selection of the AGN, 2)
comparison of different AGN types, 3) comparison of different AGN
luminosities, and 4) comparison of amplitudes derived for different
power-law slopes.  The first two concerns are strongly coupled.  Both
{\em soft} X-ray selection and optical selection preferentially detect
the unobscured, type 1 AGN.  Conversely, {\em hard} X-ray selection
also recovers the obscured, type 2 AGN.  Assuming the unified paradigm
for AGN is true (i.e., type 1 versus 2 properties are mainly due to
viewing geometry), there is no obvious reason to expect the two
classes to cluster differently.  This anticipation is supported by the
tests of \citet{Gilli2004} in the CDFN/S, though the strong angular
clustering of hard X-ray sources claimed by \citet{Yang2003} and
\citet{Basilakos2004} could be weak counter examples.  Note that the
veracity of AGN unification is debated, and arguments can be made for
the independent evolution of the two AGN populations \citep[e.g.,][and
references therein]{Franceschini2002}. Except for the
deviant CDFS result, X-ray and optical AGN cluster similarly at
similar redshifts based on the data in \mbox{Figure \ref{fig:r0evol}}.

The third issue of comparing dissimilar luminosities is relevant if
this property is correlated with the mass of the dark matter halo in
which AGN reside.  Many popular clustering models make this assumption
though its validity is not yet observationally confirmed
\citep[e.g.,][]{Croom2002}.  First consider the median luminosities of
the X-ray selected samples.  The {\em ROSAT\/} NEP AGN, with a median
luminosity of \mbox{$L_{\rm X} = 9.2 \times 10^{43}~h_{70}^{-2}$
\lxunits}, are on average about fifteen times more luminous than those
from the CDFN/S \mbox{($L_{\rm X} = 6.0 \times 10^{42}~h_{70}^{-2}$
\lxunits)}.  We estimate the median luminosity of the RASS AGN of
\citet{Akylas2000} to be a few times 10$^{43}$, thus intermediate to
the NEP and CDFN/S.  It is interesting to note that the median X-ray
luminosity of the {\em ROSAT\/} NEP AGN is rather similar to those
that we estimate for the 2QZ and AERQS.  Taking the mean absolute
magnitudes of these two samples (2QZ: $M_{B}=-25.11$,
\citealt{Croom2002}, AERQS: $M_{B}=-22.99$, \citealt{Grazian2004},
both $H_{0}=70$ km s$^{-1}$ Mpc$^{-1}$, $\Lambda$CDM) and using an
average spectral energy distribution \citep[e.g.,][]{Elvis1994} or the
observed X-ray/optical relation \citep[e.g.,][]{Vignali2003}, we find
mean luminosities of \mbox{$L_{\rm X} \approx 4 \times
10^{44}~h_{70}^{-2}$ \lxunits} for the 2QZ and \mbox{$L_{\rm X}
\approx 1 \times 10^{44}~h_{70}^{-2}$ \lxunits} for the AERQS.

The last complication to consider in regards to the data in {\mbox
Figure \ref{fig:r0evol}} lies in the fact that different authors use
different values for the power-law slope of the correlation function
in deriving the clustering amplitude.  The NEP and RASS analyses, for
example, are based on a fixed slope of $\gamma=1.8$.  The 2QZ and
AERQS adopt a slope of $\gamma=1.56$.  The CDFN/S report
single-parameter fits for their preferred value of $\gamma=1.4$.  Thus
problems may arise in comparing the clustering strengths since the
scale length and the slope are correlated to some degree.  However,
the magnitude of the effect appears to be manageable.  For instance,
in the case of the NEP these parameters are not strongly linked (see
\mbox{Figure \ref{fig:ml}}).  The NEP correlation length decreases by
$\sim$14\% to $r_{0} \approx 6.5~h^{-1}$ Mpc if we lower the slope to
$\gamma = 1.56$.  Similarly, \citet{Gilli2004} report their best-fit
correlation lengths for the CDFN/S each increase by 15\% if they
assume a slope of $\gamma=1.8$.  Notice such adjustments to homogenize
the surveys tend to erase the marginal evolution trend discussed
above, and further strengthen the case for no redshift evolution.

The above caveats notwithstanding, we briefly outline possible
theoretical interpretations of the clustering strength behavior as a
function of redshift.  The manner in which {\em mass} clusters via
gravitational instability is generally well understood.  Thus the
spatial correlation function of matter fluctuations, $\xi_{\rm mass}$,
can be computed with little ambiguity \citep[e.g.,][]{Smith2003}.
However, connecting this mass clustering to AGN clustering is
non-trivial since we do not know the distribution of AGN relative to
the distribution of mass.  This bias between the luminous and dark
matter densities is parameterized by $b(z)$ and reflects the unknown
physical mechanisms of AGN formation.

The evolution of the AGN correlation function is often formulated in
the following terms,

\begin{equation}
\xi(r,z) = b^2(z)~D^2(z)~\xi_{\rm mass}(r,z=0).
\label{biasmodel}
\end{equation}

\noindent Here $D(z)$ is the linear growth of density perturbations
which is $(1+z)^{-1}$ for an EdS cosmology.  The present-day spatial
correlation function of mass is $\xi_{\rm mass}(r,z=0)$.

We introduce three models for consideration. 
The simplest scenario of clustering evolution assumes the
bias does not change with redshift.  The linear theory model scales
like $D^2(z)$ and is represented by the solid line in \mbox{Figure
\ref{fig:r0evol}}. The second model, the long-lived scenario, proposes
that AGN are formed at an arbitrarily high redshift and then move
thereafter according to the gravitational potential of the density
fluctuations \citep{Fry1996}.  Here the bias parameter increases with
redshift which counteracts the linear growth factor and produces
negative evolution which is less steep than linear theory (dotted line
in \mbox{Figure \ref{fig:r0evol}}).  The third biasing scenario, known
as the merger model, uses the Press-Schechter (\citeyear{Press1974})
formalism to track the evolution of the dark matter halo mass function
\citep{Matarrese1997,Moscardini1998}.  The resulting bias parameter
evolves more steeply than the growth of perturbations enabling a model
clustering strength that increases with redshift.  The overall
strength of clustering reflects the mass of the dark matter halos
hosting the AGN.  The merger model predictions for several minimum
halo masses are plotted in \mbox{Figure \ref{fig:r0evol}} (dashed
lines).} More extensive discussions of these biasing models are laid
out in \citet{Croom2001} and \citet{Grazian2004}.

The linear model is clearly too steep and inconsistent with the flat
to mildly negative evolution delineated by the clustering
observations.  The less steep long-lived model agrees with the local
measures but can not replicate the flat behavior of the 2QZ data at
high redshift.  In terms of overall shape, the closest agreement is
given by the merger model.  The empirical biasing description of
\citet{Croom2001} is based on the merger model.  Their best fit with a
minimum halo mass of \mbox{$\sim$$10^{13}~h^{-1}$ M$_{\sun}$} is shown
as the dot-dashed line in \mbox{Figure \ref{fig:r0evol}}.  The {\em
ROSAT\/} NEP clustering strength is consistent with this model as are
most of the other data, except for the CDFS.

\section{Conclusions}
\label{conclusions}

We have made the first direct measurement of the spatial correlation
of X-ray selected AGN.  These results are based on the analysis of 219
AGN detected at soft energies (0.1--2.4 keV) in the {\em ROSAT\/} NEP
Survey.  The AGN catalog, presented here, includes updated X-ray and
optical properties and features complete optical identifications and
spectroscopic redshifts.  The clustering signal is significant at the
4$\sigma$ level corresponding to a clustering length of \mbox{$r_{0} =
7.4 ^{+1.8}_{-1.9}\,h^{-1}$ Mpc} assuming a power-law shape with a slope
$\gamma=1.8$.  The median redshift of the AGN contributing to this
signal is $z_{\xi}=0.22$. 

Our results indicate that X-ray and optically selected AGN share
similar clustering properties as both display clustering scale lengths
of $r_{0} \sim 6~h^{-1}$ Mpc.  This is not too surprising given that
soft X-ray and optical selection are preferentially sensitive to
unobscured type 1 AGN.  Assuming that we are probing the same
population, the {\em ROSAT\/} NEP measure provides an effective
zero-point in the study of clustering evolution.  This low redshift
determination is quite valuable since optical surveys are largely
ineffective in this regime.  The no-evolution trend delineated by the
2QZ at high redshifts appears to extend to low redshifts based on the
NEP analysis.  We find a modest increase in clustering strength
($\sim$25\%) at $z \la 0.2$ relative to the high redshift results at
$z \sim 1.5$; however, this is only 1.4$\sigma$ significant.

Since the amplitudes of galaxy and AGN clustering are similar, we
argue that AGN are not biased relative to galaxies.  It likely that
AGN randomly sample the galaxy distribution and do not preferentially
probe the high peaks of the matter density field.  Given that the bias
of local galaxies is near unity, the inferred X-ray bias is near
unity.  Hence it seems that X-ray selected AGN closely trace the
underlying mass distribution.

The study of X-ray AGN clustering is still in the early stages and
much work lies ahead.  The NEP has provided the first firm data point,
and the recent results from the CDFN/S provide small uncertainty but
large variance measures at high redshift.  Forthcoming results from
on-going and planned deep surveys with {\em Chandra} and {\em
XMM-Newton\/} will play a vital role by hopefully averaging out cosmic
variance and securing the clustering amplitude at high redshift.
Future X-ray survey missions like the Dark Universe Observatory will
provide precision measures of X-ray AGN clustering as a function of
high redshift with the {\em ROSAT\/} NEP results serving as an
important anchor point at low redshift.\\

\acknowledgements

It is a pleasure to acknowledge Roberto Gilli, Emanuele Daddi, and
Anna Wolter for stimulating discussions.  We thank the first two for
allowing us to quote their results for the CDFN/S prior to
publication.  We appreciate Giusi Micela and Jorge Sanz for providing
additional observations to clarify the identification of
RX\,J1824.7+6509.  We thank the anonymous referee for the careful
review of this work and for comments that improved the
presentation. We are grateful to our sponsoring agencies. We are
grateful to our sponsoring agencies.  Support has come from the ESO
Office for Science, NSF (AST 91-19216 and AST \mbox{95-00515}), NASA
(NGT5-50175, GO-5402.01-93A, and GO-05987.02-94A), the ARCS
foundation, the Smithsonian Institution, NATO (CRG91-0415), the
Italian Space Agency ASI-CNR, the Bundesministerium f\"{u}r Forschung
(BMBF/DLR), and the Max-Planck-Gesellschaft (MPG).

\bibliographystyle{apj}
\bibliography{myrefs}
 
\newpage
\clearpage
\input{tab1.tex}

\end{document}

%% file: tab1.tex
\LongTables
\begin{landscape}
\begin{deluxetable}{ccccccccccccc}
\tablewidth{0pt}
\tabletypesize{\scriptsize}
\tablecaption{{\em ROSAT\/} NEP AGN Catalog\label{tab:agn}}
\tablehead{\colhead{Object} & 
           \colhead{NEP} & 
           \colhead{$\alpha_{\rm X}$} & 
           \colhead{$\delta_{\rm X}$} & 
           \colhead{$\alpha_{\rm opt}$} & 
           \colhead{$\delta_{\rm opt}$} &
           \colhead{$n_{H}$} &
           \colhead{Count Rate} & 
           \colhead{Count Rate} & 
           \colhead{$f_{\rm X}$} & 
           \colhead{$L_{\rm X}$} & 
           \colhead{$z$} & 
	   \colhead{AGN} \\
           \colhead{} & 
           \colhead{ID} & 
           \colhead{(J2000)} & 
           \colhead{(J2000)} & 
           \colhead{(J2000)} & 
           \colhead{(J2000)} &
           \colhead{[$10^{20}$ cm$^{-2}$]} &
           \colhead{[s$^{-1}$]} & 
           \colhead{Error [s$^{-1}$]} & 
           \colhead{[$10^{-14}$ cgs]} & 
           \colhead{[$10^{44}$ cgs]} & 
           \colhead{} & 
           \colhead{Type} \\
           \colhead{(1)} & 
           \colhead{(2)} & 
           \colhead{(3)} & 
           \colhead{(4)} & 
           \colhead{(5)} & 
           \colhead{(6)} &
           \colhead{(7)} &
           \colhead{(8)} & 
           \colhead{(9)} & 
           \colhead{(10)} & 
           \colhead{(11)} & 
           \colhead{(12)} & 
           \colhead{(13)} }
\startdata
\input{tab1-data.tex}
\enddata
\tablenotetext{$\star$}{RX\,J1824.7+6509 (NEP 5500) was formerly classified as a stellar X-ray source based on the literature.  However, more recent work convincingly demonstrates this source is a type 1 AGN at $z=0.303$ \citep[e.g.,][]{Engels1998}.}
\tablenotetext{}{{\sc Note} --- Table 1 is also available in machine-readable form in the eletronic edition of the {\em Astrophysical Journal}.}
\end{deluxetable}
\clearpage
\end{landscape}

%% file: tab1-data.tex
RX\,J1715.4+6239 & 1239 & 17 15 25.3 & +62 39 34 & 17 15 25.7 & +62 39 27 & 2.62 & 0.0173 & 0.0038 & \phantom{0}\phantom{0}13.94\phantom{0}&\phantom{0} $ \phantom{0}\phantom{0}4.88\phantom{0} $ & 0.8500 & 2\\
RX\,J1716.2+6836 & 1270 & 17 16 14.4 & +68 36 36 & 17 16 13.8 & +68 36 38 & 3.57 & 0.1192 & 0.0059 & \phantom{0}112.56\phantom{0}&\phantom{0} $ \phantom{0}31.57\phantom{0} $ & 0.7770 & 1\\
RX\,J1717.1+6401 & 1272 & 17 17 08.1 & +64 01 46 & 17 17 07.1 & +64 01 45 & 2.97 & 0.0400 & 0.0043 & \phantom{0}\phantom{0}34.41\phantom{0}&\phantom{0} $ \phantom{0}\phantom{0}0.16\phantom{0} $ & 0.1334 & 1\\
RX\,J1717.5+6559 & 1300 & 17 17 35.6 & +65 59 35 & 17 17 37.9 & +65 59 39 & 3.31 & 0.0467 & 0.0052 & \phantom{0}\phantom{0}42.45\phantom{0}&\phantom{0} $ \phantom{0}\phantom{0}1.17\phantom{0} $ & 0.2936 & 1\\
RX\,J1717.7+6431 & 1320 & 17 17 44.4 & +64 31 45 & 17 17 47.4 & +64 31 41 & 3.61 & 0.0164 & 0.0029 & \phantom{0}\phantom{0}15.59\phantom{0}&\phantom{0} $ \phantom{0}\phantom{0}0.004 $ & 0.0337 & 2\\
RX\,J1717.9+7038 & 1330 & 17 17 57.0 & +70 38 15 & 17 17 56.6 & +70 38 16 & 3.96 & 0.0492 & 0.0048 & \phantom{0}\phantom{0}48.91\phantom{0}&\phantom{0} $ \phantom{0}\phantom{0}0.41\phantom{0} $ & 0.1738 & 2\\
RX\,J1718.0+6727 & 1340 & 17 18 05.5 & +67 27 11 & 17 18 05.9 & +67 27 00 & 3.77 & 0.0207 & 0.0028 & \phantom{0}\phantom{0}20.09\phantom{0}&\phantom{0} $ \phantom{0}\phantom{0}2.43\phantom{0} $ & 0.5506 & 1\\
RX\,J1719.0+6929 & 1410 & 17 19 03.6 & +69 29 33 & 17 19 03.5 & +69 29 39 & 4.02 & 0.0129 & 0.0027 & \phantom{0}\phantom{0}12.93\phantom{0}&\phantom{0} $ \phantom{0}\phantom{0}0.32\phantom{0} $ & 0.2816 & 1\\
RX\,J1720.1+6833 & 1450 & 17 20 07.8 & +68 33 37 & 17 20 06.6 & +68 33 50 & 3.64 & 0.0175 & 0.0027 & \phantom{0}\phantom{0}16.71\phantom{0}&\phantom{0} $ \phantom{0}\phantom{0}1.92\phantom{0} $ & 0.5392 & 1\\
RX\,J1720.8+6210 & 1471 & 17 20 48.8 & +62 10 13 & 17 20 42.3 & +62 10 10 & 2.80 & 0.0251 & 0.0051 & \phantom{0}\phantom{0}20.92\phantom{0}&\phantom{0} $ \phantom{0}\phantom{0}5.06\phantom{0} $ & 0.7313 & 1\\
RX\,J1721.0+6711 & 1490 & 17 21 03.1 & +67 11 54 & 17 21 02.3 & +67 11 57 & 3.77 & 0.0099 & 0.0021 & \phantom{0}\phantom{0}\phantom{0}9.61\phantom{0}&\phantom{0} $ \phantom{0}\phantom{0}2.50\phantom{0} $ & 0.7538 & 1\\
RX\,J1723.1+6826 & 1540 & 17 23 10.4 & +68 26 54 & 17 23 09.9 & +68 26 56 & 4.18 & 0.0082 & 0.0019 & \phantom{0}\phantom{0}\phantom{0}8.37\phantom{0}&\phantom{0} $ \phantom{0}\phantom{0}4.14\phantom{0} $ & 0.9782 & 1\\
RX\,J1724.9+6636 & 1640 & 17 24 55.5 & +66 36 59 & 17 24 56.1 & +66 36 50 & 3.84 & 0.0105 & 0.0026 & \phantom{0}\phantom{0}10.29\phantom{0}&\phantom{0} $ \phantom{0}\phantom{0}2.08\phantom{0} $ & 0.6792 & 2\\
RX\,J1726.5+6714 & 1670 & 17 26 30.7 & +67 14 12 & 17 26 28.6 & +67 14 15 & 3.91 & 0.0108 & 0.0020 & \phantom{0}\phantom{0}10.67\phantom{0}&\phantom{0} $ \phantom{0}\phantom{0}0.49\phantom{0} $ & 0.3659 & 1\\
RX\,J1726.7+6643 & 1680 & 17 26 43.9 & +66 43 30 & 17 26 45.0 & +66 43 19 & 3.73 & 0.0399 & 0.0044 & \phantom{0}\phantom{0}38.55\phantom{0}&\phantom{0} $ \phantom{0}\phantom{0}1.82\phantom{0} $ & 0.3705 & 1\\
RX\,J1727.2+6322 & 1710 & 17 27 12.0 & +63 22 44 & 17 27 11.7 & +63 22 41 & 2.88 & 0.0818 & 0.0070 & \phantom{0}\phantom{0}69.20\phantom{0}&\phantom{0} $ \phantom{0}\phantom{0}0.95\phantom{0} $ & 0.2169 & 1\\
RX\,J1727.8+6748 & 1740 & 17 27 49.5 & +67 48 43 & 17 27 45.5 & +67 48 43 & 4.16 & 0.0142 & 0.0021 & \phantom{0}\phantom{0}14.47\phantom{0}&\phantom{0} $ \phantom{0}\phantom{0}1.36\phantom{0} $ & 0.4950 & 1\\
RX\,J1728.5+6732 & 1770 & 17 28 35.4 & +67 32 33 & 17 28 34.6 & +67 32 24 & 4.56 & 0.0071 & 0.0016 & \phantom{0}\phantom{0}\phantom{0}7.55\phantom{0}&\phantom{0} $ \phantom{0}\phantom{0}1.36\phantom{0} $ & 0.6493 & 1\\
RX\,J1729.2+7032 & 1782 & 17 29 12.0 & +70 32 57 & 17 29 11.8 & +70 32 55 & 3.88 & 0.0140 & 0.0032 & \phantom{0}\phantom{0}13.78\phantom{0}&\phantom{0} $ \phantom{0}\phantom{0}1.58\phantom{0} $ & 0.5378 & 1\\
RX\,J1732.0+6926 & 1910 & 17 32 05.5 & +69 26 22 & 17 32 04.5 & +69 26 39 & 4.02 & 0.0148 & 0.0026 & \phantom{0}\phantom{0}14.83\phantom{0}&\phantom{0} $ \phantom{0}\phantom{0}1.45\phantom{0} $ & 0.5043 & 1\\
RX\,J1732.5+7031 & 1920 & 17 32 31.3 & +70 31 37 & 17 32 31.0 & +70 31 31 & 3.88 & 0.0363 & 0.0045 & \phantom{0}\phantom{0}35.73\phantom{0}&\phantom{0} $ \phantom{0}\phantom{0}0.47\phantom{0} $ & 0.2114 & 1\\
RX\,J1732.9+6533 & 1930 & 17 32 54.5 & +65 33 24 & 17 32 53.9 & +65 33 25 & 4.01 & 0.0344 & 0.0034 & \phantom{0}\phantom{0}34.42\phantom{0}&\phantom{0} $ \phantom{0}12.25\phantom{0} $ & 0.8560 & 1\\
RX\,J1734.5+6755 & 1980 & 17 34 30.3 & +67 55 05 & 17 34 27.8 & +67 55 04 & 4.83 & 0.0080 & 0.0017 & \phantom{0}\phantom{0}\phantom{0}8.73\phantom{0}&\phantom{0} $ \phantom{0}\phantom{0}0.14\phantom{0} $ & 0.2341 & 1\\
RX\,J1736.0+6559 & 2040 & 17 36 00.0 & +65 59 00 & 17 36 01.9 & +65 58 54 & 3.74 & 0.0094 & 0.0021 & \phantom{0}\phantom{0}\phantom{0}9.09\phantom{0}&\phantom{0} $ \phantom{0}\phantom{0}0.62\phantom{0} $ & 0.4341 & 1\\
RX\,J1737.0+6601 & 2131 & 17 37 05.5 & +66 01 05 & 17 37 07.8 & +66 01 02 & 3.76 & 0.0141 & 0.0021 & \phantom{0}\phantom{0}13.68\phantom{0}&\phantom{0} $ \phantom{0}\phantom{0}0.60\phantom{0} $ & 0.3580 & 2\\
RX\,J1738.0+6210 & 2160 & 17 38 02.6 & +62 10 42 & 17 38 03.8 & +62 10 54 & 3.34 & 0.0115 & 0.0024 & \phantom{0}\phantom{0}10.51\phantom{0}&\phantom{0} $ \phantom{0}13.54\phantom{0} $ & 1.4402 & 1\\
RX\,J1738.4+6417 & 2180 & 17 38 24.0 & +64 17 59 & 17 38 24.5 & +64 17 56 & 2.77 & 0.0131 & 0.0024 & \phantom{0}\phantom{0}10.87\phantom{0}&\phantom{0} $ \phantom{0}\phantom{0}3.23\phantom{0} $ & 0.7955 & 1\\
RX\,J1738.7+7037 & 2200 & 17 38 42.0 & +70 37 05 & 17 38 42.0 & +70 37 16 & 3.93 & 0.0289 & 0.0037 & \phantom{0}\phantom{0}28.64\phantom{0}&\phantom{0} $ \phantom{0}\phantom{0}0.15\phantom{0} $ & 0.1399 & 1\\
RX\,J1739.3+6614 & 2211 & 17 39 21.5 & +66 14 41 & 17 39 25.8 & +66 14 31 & 3.66 & 0.0078 & 0.0019 & \phantom{0}\phantom{0}\phantom{0}7.46\phantom{0}&\phantom{0} $ \phantom{0}\phantom{0}8.13\phantom{0} $ & 1.3460 & 1\\
RX\,J1739.7+6710 & 2230 & 17 39 44.6 & +67 10 52 & 17 39 44.7 & +67 10 43 & 4.49 & 0.0339 & 0.0027 & \phantom{0}\phantom{0}35.79\phantom{0}&\phantom{0} $ \phantom{0}\phantom{0}0.13\phantom{0} $ & 0.1180 & 1\\
RX\,J1739.9+7005 & 2240 & 17 39 56.0 & +70 05 52 & 17 39 54.2 & +70 05 57 & 3.93 & 0.0169 & 0.0028 & \phantom{0}\phantom{0}16.75\phantom{0}&\phantom{0} $ \phantom{0}\phantom{0}1.78\phantom{0} $ & 0.5209 & 1\\
RX\,J1741.2+6507 & 2300 & 17 41 14.4 & +65 07 43 & 17 41 15.7 & +65 07 42 & 4.26 & 0.0133 & 0.0021 & \phantom{0}\phantom{0}13.71\phantom{0}&\phantom{0} $ \phantom{0}\phantom{0}3.49\phantom{0} $ & 0.7466 & 1\\
RX\,J1741.7+6335 & 2320 & 17 41 46.0 & +63 35 13 & 17 41 45.6 & +63 35 22 & 2.75 & 0.0169 & 0.0026 & \phantom{0}\phantom{0}13.95\phantom{0}&\phantom{0} $ \phantom{0}65.61\phantom{0} $ & 2.4420 & 1\\
RX\,J1742.2+6639 & 2340 & 17 42 12.5 & +66 39 49 & 17 42 13.8 & +66 39 34 & 3.69 & 0.0132 & 0.0020 & \phantom{0}\phantom{0}12.69\phantom{0}&\phantom{0} $ \phantom{0}12.02\phantom{0} $ & 1.2720 & 1\\
RX\,J1742.2+6936 & 2341 & 17 42 15.1 & +69 36 29 & 17 42 16.6 & +69 36 21 & 3.71 & 0.0100 & 0.0023 & \phantom{0}\phantom{0}\phantom{0}9.63\phantom{0}&\phantom{0} $ \phantom{0}\phantom{0}5.64\phantom{0} $ & 1.0470 & 1\\
RX\,J1742.2+6351 & 2350 & 17 42 17.9 & +63 51 09 & 17 42 18.4 & +63 51 15 & 2.91 & 0.0389 & 0.0038 & \phantom{0}\phantom{0}33.13\phantom{0}&\phantom{0} $ \phantom{0}\phantom{0}1.89\phantom{0} $ & 0.4019 & 1\\
RX\,J1742.7+6800 & 2371 & 17 42 42.0 & +68 00 11 & 17 42 43.5 & +68 00 16 & 3.99 & 0.0084 & 0.0016 & \phantom{0}\phantom{0}\phantom{0}8.38\phantom{0}&\phantom{0} $ \phantom{0}\phantom{0}0.02\phantom{0} $ & 0.0858 & 1\\
RX\,J1743.7+6829 & 2450 & 17 43 43.5 & +68 29 26 & 17 43 42.9 & +68 29 25 & 4.26 & 0.0066 & 0.0015 & \phantom{0}\phantom{0}\phantom{0}6.80\phantom{0}&\phantom{0} $ \phantom{0}\phantom{0}0.28\phantom{0} $ & 0.3504 & 1\\
RX\,J1743.8+6657 & 2460 & 17 43 49.2 & +66 57 23 & 17 43 49.3 & +66 57 08 & 4.13 & 0.0055 & 0.0013 & \phantom{0}\phantom{0}\phantom{0}5.58\phantom{0}&\phantom{0} $ \phantom{0}\phantom{0}0.51\phantom{0} $ & 0.4900 & 1\\
RX\,J1744.2+6534 & 2490 & 17 44 14.2 & +65 34 54 & 17 44 14.5 & +65 34 53 & 3.58 & 0.0544 & 0.0030 & \phantom{0}\phantom{0}51.50\phantom{0}&\phantom{0} $ \phantom{0}\phantom{0}1.02\phantom{0} $ & 0.2550 & 1\\
RX\,J1744.9+6536 & 2550 & 17 44 55.0 & +65 36 00 & 17 44 54.5 & +65 36 02 & 3.59 & 0.0061 & 0.0012 & \phantom{0}\phantom{0}\phantom{0}5.78\phantom{0}&\phantom{0} $ \phantom{0}\phantom{0}0.24\phantom{0} $ & 0.3533 & 1\\
RX\,J1745.7+6748 & 2610 & 17 45 42.6 & +67 48 15 & 17 45 42.4 & +67 48 14 & 4.28 & 0.0058 & 0.0013 & \phantom{0}\phantom{0}\phantom{0}5.99\phantom{0}&\phantom{0} $ \phantom{0}\phantom{0}0.37\phantom{0} $ & 0.4143 & 2\\
RX\,J1745.9+6451 & 2650 & 17 45 55.2 & +64 51 18 & 17 45 55.5 & +64 51 25 & 3.49 & 0.0211 & 0.0024 & \phantom{0}\phantom{0}19.70\phantom{0}&\phantom{0} $ \phantom{0}\phantom{0}0.18\phantom{0} $ & 0.1790 & 1\\
RX\,J1746.0+6727 & 2670 & 17 46 03.0 & +67 27 09 & 17 46 01.8 & +67 27 09 & 4.70 & 0.0120 & 0.0015 & \phantom{0}\phantom{0}12.94\phantom{0}&\phantom{0} $ \phantom{0}\phantom{0}0.17\phantom{0} $ & 0.2146 & 1\\
RX\,J1746.1+6737 & 2700 & 17 46 09.6 & +67 37 21 & 17 46 08.8 & +67 37 15 & 4.21 & 0.1464 & 0.0044 & \phantom{0}149.98\phantom{0}&\phantom{0} $ \phantom{0}\phantom{0}0.06\phantom{0} $ & 0.0410 & 1\\
RX\,J1746.2+6227 & 2710 & 17 46 14.6 & +62 27 01 & 17 46 13.9 & +62 26 54 & 3.25 & 0.0319 & 0.0034 & \phantom{0}\phantom{0}28.73\phantom{0}&\phantom{0} $ 413.35\phantom{0} $ & 3.8890 & 1\\
RX\,J1746.3+6320 & 2750 & 17 46 21.6 & +63 20 06 & 17 46 21.8 & +63 20 10 & 3.02 & 0.0165 & 0.0026 & \phantom{0}\phantom{0}14.31\phantom{0}&\phantom{0} $ \phantom{0}\phantom{0}0.67\phantom{0} $ & 0.3697 & 1\\
RX\,J1747.0+6836 & 2800 & 17 47 00.3 & +68 36 26 & 17 46 59.9 & +68 36 34 & 4.42 & 0.2131 & 0.0060 & \phantom{0}223.35\phantom{0}&\phantom{0} $ \phantom{0}\phantom{0}0.21\phantom{0} $ & 0.0630 & 1\\
RX\,J1747.1+6813 & 2810 & 17 47 10.6 & +68 13 19 & 17 47 12.7 & +68 13 26 & 4.54 & 0.0057 & 0.0013 & \phantom{0}\phantom{0}\phantom{0}6.05\phantom{0}&\phantom{0} $ \phantom{0}27.06\phantom{0} $ & 2.3920 & 1\\
RX\,J1747.2+6532 & 2820 & 17 47 14.4 & +65 32 30 & 17 47 13.9 & +65 32 35 & 3.66 & 0.0108 & 0.0016 & \phantom{0}\phantom{0}10.33\phantom{0}&\phantom{0} $ \phantom{0}15.11\phantom{0} $ & 1.5166 & 1\\
RX\,J1747.3+6702 & 2840 & 17 47 22.2 & +67 02 06 & 17 47 21.5 & +67 02 01 & 4.51 & 0.0066 & 0.0011 & \phantom{0}\phantom{0}\phantom{0}6.99\phantom{0}&\phantom{0} $ \phantom{0}\phantom{0}1.75\phantom{0} $ & 0.7421 & 1\\
RX\,J1747.4+6626 & 2850 & 17 47 26.8 & +66 26 27 & 17 47 27.0 & +66 26 24 & 3.78 & 0.0138 & 0.0014 & \phantom{0}\phantom{0}13.42\phantom{0}&\phantom{0} $ \phantom{0}\phantom{0}0.07\phantom{0} $ & 0.1391 & 1\\
RX\,J1747.4+6924 & 2860 & 17 47 27.0 & +69 24 55 & 17 47 27.9 & +69 25 09 & 3.49 & 0.0099 & 0.0020 & \phantom{0}\phantom{0}\phantom{0}9.25\phantom{0}&\phantom{0} $ \phantom{0}\phantom{0}1.02\phantom{0} $ & 0.5292 & 2\\
RX\,J1747.9+6538 & 2890 & 17 47 58.0 & +65 38 35 & 17 47 57.9 & +65 38 28 & 3.85 & 0.0305 & 0.0022 & \phantom{0}\phantom{0}29.91\phantom{0}&\phantom{0} $ \phantom{0}\phantom{0}1.04\phantom{0} $ & 0.3248 & 1\\
RX\,J1748.2+7016 & 2900 & 17 48 17.4 & +70 16 14 & 17 48 19.6 & +70 16 09 & 3.93 & 0.0446 & 0.0036 & \phantom{0}\phantom{0}44.20\phantom{0}&\phantom{0} $ \phantom{0}\phantom{0}0.43\phantom{0} $ & 0.1858 & 1\\
RX\,J1748.3+6403 & 2910 & 17 48 22.7 & +64 03 27 & 17 48 23.1 & +64 03 38 & 3.27 & 0.0155 & 0.0023 & \phantom{0}\phantom{0}14.00\phantom{0}&\phantom{0} $ \phantom{0}\phantom{0}7.06\phantom{0} $ & 0.9859 & 1\\
RX\,J1748.6+6842 & 2940 & 17 48 38.8 & +68 42 11 & 17 48 38.3 & +68 42 17 & 4.25 & 0.0236 & 0.0023 & \phantom{0}\phantom{0}24.28\phantom{0}&\phantom{0} $ \phantom{0}\phantom{0}0.02\phantom{0} $ & 0.0537 & 1\\
RX\,J1749.3+6411 & 3000 & 17 49 20.4 & +64 11 08 & 17 49 19.5 & +64 11 19 & 3.27 & 0.0109 & 0.0019 & \phantom{0}\phantom{0}\phantom{0}9.85\phantom{0}&\phantom{0} $ \phantom{0}\phantom{0}4.94\phantom{0} $ & 0.9836 & 1\\
RX\,J1749.7+6422 & 3001 & 17 49 42.4 & +64 22 46 & 17 49 44.1 & +64 22 58 & 3.24 & 0.0078 & 0.0017 & \phantom{0}\phantom{0}\phantom{0}7.02\phantom{0}&\phantom{0} $ \phantom{0}\phantom{0}1.83\phantom{0} $ & 0.7540 & 1\\
RX\,J1750.2+6814 & 3050 & 17 50 14.3 & +68 14 33 & 17 50 16.1 & +68 14 37 & 4.50 & 0.0069 & 0.0013 & \phantom{0}\phantom{0}\phantom{0}7.29\phantom{0}&\phantom{0} $ \phantom{0}\phantom{0}0.12\phantom{0} $ & 0.2310 & 1\\
RX\,J1750.2+6415 & 3060 & 17 50 15.5 & +64 15 15 & 17 50 15.1 & +64 14 56 & 3.11 & 0.0115 & 0.0019 & \phantom{0}\phantom{0}10.13\phantom{0}&\phantom{0} $ \phantom{0}\phantom{0}0.19\phantom{0} $ & 0.2504 & 2\\
RX\,J1751.0+6710 & 3100 & 17 51 02.4 & +67 10 09 & 17 51 01.2 & +67 10 14 & 4.33 & 0.0038 & 0.0008 & \phantom{0}\phantom{0}\phantom{0}3.95\phantom{0}&\phantom{0} $ \phantom{0}\phantom{0}0.56\phantom{0} $ & 0.5870 & 1\\
RX\,J1751.1+6753 & 3120 & 17 51 09.5 & +67 53 07 & 17 51 08.9 & +67 53 08 & 4.45 & 0.0062 & 0.0011 & \phantom{0}\phantom{0}\phantom{0}6.52\phantom{0}&\phantom{0} $ \phantom{0}\phantom{0}0.25\phantom{0} $ & 0.3406 & 1\\
RX\,J1751.6+6540 & 3160 & 17 51 39.7 & +65 40 40 & 17 51 36.9 & +65 40 30 & 4.10 & 0.0122 & 0.0015 & \phantom{0}\phantom{0}12.33\phantom{0}&\phantom{0} $ \phantom{0}\phantom{0}4.02\phantom{0} $ & 0.8259 & 1\\
RX\,J1751.9+6551 & 3190 & 17 51 57.6 & +65 51 20 & 17 51 56.7 & +65 51 17 & 4.13 & 0.0121 & 0.0012 & \phantom{0}\phantom{0}12.28\phantom{0}&\phantom{0} $ \phantom{0}\phantom{0}0.65\phantom{0} $ & 0.3901 & 1\\
RX\,J1752.2+6624 & 3210 & 17 52 12.6 & +66 24 56 & 17 52 11.7 & +66 24 54 & 3.88 & 0.0049 & 0.0008 & \phantom{0}\phantom{0}\phantom{0}4.82\phantom{0}&\phantom{0} $ \phantom{0}\phantom{0}0.27\phantom{0} $ & 0.4002 & 1\\
RX\,J1752.9+6440 & 3230 & 17 52 57.4 & +64 40 58 & 17 52 56.9 & +64 40 56 & 3.22 & 0.0125 & 0.0016 & \phantom{0}\phantom{0}11.21\phantom{0}&\phantom{0} $ \phantom{0}\phantom{0}0.04\phantom{0} $ & 0.1230 & 1\\
RX\,J1753.1+6746 & \phantom{0}\phantom{0}40 & 17 53 09.7 & +67 46 44 & 17 53 09.6 & +67 46 32 & 4.77 & 0.0070 & 0.0011 & \phantom{0}\phantom{0}\phantom{0}7.60\phantom{0}&\phantom{0} $ \phantom{0}\phantom{0}5.37\phantom{0} $ & 1.1297 & 1\\
RX\,J1753.5+6811 & 3231 & 17 53 30.9 & +68 11 47 & 17 53 32.4 & +68 12 01 & 5.01 & 0.0056 & 0.0014 & \phantom{0}\phantom{0}\phantom{0}6.21\phantom{0}&\phantom{0} $ \phantom{0}\phantom{0}0.43\phantom{0} $ & 0.4366 & 2\\
RX\,J1753.6+6542 & 3260 & 17 53 41.6 & +65 42 42 & 17 53 42.1 & +65 42 40 & 4.34 & 0.0076 & 0.0009 & \phantom{0}\phantom{0}\phantom{0}7.90\phantom{0}&\phantom{0} $ \phantom{0}\phantom{0}0.04\phantom{0} $ & 0.1400 & 1\\
RX\,J1753.9+7016 & 3280 & 17 53 55.5 & +70 16 47 & 17 53 56.6 & +70 16 42 & 4.25 & 0.0095 & 0.0018 & \phantom{0}\phantom{0}\phantom{0}9.77\phantom{0}&\phantom{0} $ \phantom{0}\phantom{0}0.01\phantom{0} $ & 0.0620 & 1\\
RX\,J1754.0+6613 & \phantom{0}\phantom{0}60 & 17 54 05.4 & +66 13 54 & 17 54 04.8 & +66 13 50 & 4.12 & 0.0152 & 0.0010 & \phantom{0}\phantom{0}15.40\phantom{0}&\phantom{0} $ \phantom{0}\phantom{0}0.90\phantom{0} $ & 0.4067 & 1\\
RX\,J1754.7+6819 & 3340 & 17 54 42.3 & +68 19 08 & 17 54 42.0 & +68 19 06 & 5.23 & 0.0066 & 0.0011 & \phantom{0}\phantom{0}\phantom{0}7.46\phantom{0}&\phantom{0} $ \phantom{0}\phantom{0}0.27\phantom{0} $ & 0.3292 & 1\\
RX\,J1754.7+6208 & 3350 & 17 54 43.2 & +62 08 21 & 17 54 42.3 & +62 08 30 & 3.41 & 0.0147 & 0.0024 & \phantom{0}\phantom{0}13.58\phantom{0}&\phantom{0} $ \phantom{0}\phantom{0}0.45\phantom{0} $ & 0.3190 & 1\\
RX\,J1754.8+6706 & \phantom{0}100 & 17 54 49.3 & +67 06 00 & 17 54 49.7 & +67 05 56 & 4.53 & 0.0023 & 0.0005 & \phantom{0}\phantom{0}\phantom{0}2.44\phantom{0}&\phantom{0} $ \phantom{0}\phantom{0}1.03\phantom{0} $ & 0.9190 & 1\\
RX\,J1755.0+6446 & 3360 & 17 55 00.0 & +64 46 32 & 17 55 00.8 & +64 46 32 & 3.49 & 0.0085 & 0.0013 & \phantom{0}\phantom{0}\phantom{0}7.94\phantom{0}&\phantom{0} $ \phantom{0}\phantom{0}1.65\phantom{0} $ & 0.6870 & 1\\
RX\,J1755.0+6235 & 3361 & 17 55 03.6 & +62 35 30 & 17 55 03.2 & +62 35 41 & 3.33 & 0.0083 & 0.0020 & \phantom{0}\phantom{0}\phantom{0}7.57\phantom{0}&\phantom{0} $ \phantom{0}13.86\phantom{0} $ & 1.6607 & 1\\
RX\,J1755.0+6519 & 3370 & 17 55 05.8 & +65 19 50 & 17 55 05.6 & +65 19 55 & 3.78 & 0.0948 & 0.0030 & \phantom{0}\phantom{0}92.17\phantom{0}&\phantom{0} $ \phantom{0}\phantom{0}0.14\phantom{0} $ & 0.0785 & 1\\
RX\,J1755.1+6719 & \phantom{0}110 & 17 55 09.0 & +67 19 50 & 17 55 08.3 & +67 19 54 & 4.69 & 0.0049 & 0.0008 & \phantom{0}\phantom{0}\phantom{0}5.28\phantom{0}&\phantom{0} $ \phantom{0}\phantom{0}0.08\phantom{0} $ & 0.2225 & 2\\
RX\,J1755.1+6852 & 3380 & 17 55 11.9 & +68 52 30 & 17 55 10.7 & +68 52 34 & 4.56 & 0.0062 & 0.0012 & \phantom{0}\phantom{0}\phantom{0}6.59\phantom{0}&\phantom{0} $ \phantom{0}\phantom{0}7.43\phantom{0} $ & 1.3645 & 1\\
RX\,J1755.6+6209 & 3420 & 17 55 40.3 & +62 09 41 & 17 55 40.3 & +62 09 39 & 3.41 & 0.0240 & 0.0027 & \phantom{0}\phantom{0}22.17\phantom{0}&\phantom{0} $ \phantom{0}\phantom{0}0.04\phantom{0} $ & 0.0846 & 2\\
RX\,J1755.6+7009 & 3410 & 17 55 40.5 & +70 09 52 & 17 55 41.3 & +70 09 51 & 4.25 & 0.0077 & 0.0016 & \phantom{0}\phantom{0}\phantom{0}7.92\phantom{0}&\phantom{0} $ \phantom{0}\phantom{0}0.53\phantom{0} $ & 0.4295 & 1\\
RX\,J1755.7+6249 & 3430 & 17 55 46.2 & +62 49 27 & 17 55 45.9 & +62 49 29 & 3.33 & 0.0402 & 0.0034 & \phantom{0}\phantom{0}36.67\phantom{0}&\phantom{0} $ \phantom{0}\phantom{0}0.61\phantom{0} $ & 0.2360 & 1\\
RX\,J1755.9+6540 & \phantom{0}140 & 17 55 56.9 & +65 40 54 & 17 55 56.8 & +65 40 52 & 4.15 & 0.0089 & 0.0009 & \phantom{0}\phantom{0}\phantom{0}9.05\phantom{0}&\phantom{0} $ \phantom{0}\phantom{0}0.31\phantom{0} $ & 0.3238 & 1\\
RX\,J1756.1+6615 & \phantom{0}160 & 17 56 10.0 & +66 15 14 & 17 56 09.5 & +66 15 09 & 4.01 & 0.0086 & 0.0007 & \phantom{0}\phantom{0}\phantom{0}8.60\phantom{0}&\phantom{0} $ \phantom{0}\phantom{0}1.48\phantom{0} $ & 0.6357 & 1\\
RX\,J1756.1+7001 & 3460 & 17 56 10.8 & +70 01 55 & 17 56 11.9 & +70 01 47 & 3.90 & 0.0090 & 0.0017 & \phantom{0}\phantom{0}\phantom{0}8.88\phantom{0}&\phantom{0} $ \phantom{0}\phantom{0}0.12\phantom{0} $ & 0.2129 & 1\\
RX\,J1756.1+7055 & 3470 & 17 56 10.8 & +70 55 48 & 17 56 11.6 & +70 55 50 & 4.25 & 0.0163 & 0.0022 & \phantom{0}\phantom{0}16.76\phantom{0}&\phantom{0} $ \phantom{0}\phantom{0}0.31\phantom{0} $ & 0.2460 & 1\\
RX\,J1756.2+6619 & \phantom{0}170 & 17 56 12.0 & +66 19 47 & 17 56 12.2 & +66 19 46 & 4.01 & 0.0023 & 0.0004 & \phantom{0}\phantom{0}\phantom{0}2.30\phantom{0}&\phantom{0} $ \phantom{0}\phantom{0}1.64\phantom{0} $ & 1.1340 & 1\\
RX\,J1756.2+6955 & 3480 & 17 56 12.7 & +69 55 21 & 17 56 12.7 & +69 55 20 & 3.84 & 0.0177 & 0.0023 & \phantom{0}\phantom{0}17.33\phantom{0}&\phantom{0} $ \phantom{0}\phantom{0}0.03\phantom{0} $ & 0.0838 & 2\\
RX\,J1756.2+6838 & 3490 & 17 56 13.6 & +68 38 31 & 17 56 15.5 & +68 38 25 & 5.24 & 0.0070 & 0.0012 & \phantom{0}\phantom{0}\phantom{0}7.92\phantom{0}&\phantom{0} $ \phantom{0}\phantom{0}0.02\phantom{0} $ & 0.1019 & 2\\
RX\,J1756.4+6300 & 3510 & 17 56 25.2 & +63 00 42 & 17 56 25.4 & +63 00 49 & 3.53 & 0.0077 & 0.0016 & \phantom{0}\phantom{0}\phantom{0}7.23\phantom{0}&\phantom{0} $ \phantom{0}\phantom{0}4.90\phantom{0} $ & 1.1110 & 1\\
RX\,J1756.7+6438 & 3530 & 17 56 43.2 & +64 38 53 & 17 56 43.4 & +64 38 59 & 3.49 & 0.0231 & 0.0019 & \phantom{0}\phantom{0}21.57\phantom{0}&\phantom{0} $ \phantom{0}\phantom{0}0.32\phantom{0} $ & 0.2233 & 1\\
RX\,J1756.8+6612 & \phantom{0}180 & 17 56 52.4 & +66 12 42 & 17 56 51.3 & +66 12 42 & 4.14 & 0.0061 & 0.0006 & \phantom{0}\phantom{0}\phantom{0}6.20\phantom{0}&\phantom{0} $ \phantom{0}\phantom{0}7.78\phantom{0} $ & 1.4252 & 1\\
RX\,J1756.9+6238 & 3550 & 17 56 58.2 & +62 38 44 & 17 57 00.6 & +62 38 55 & 3.33 & 0.0205 & 0.0026 & \phantom{0}\phantom{0}18.70\phantom{0}&\phantom{0} $ \phantom{0}12.10\phantom{0} $ & 1.0902 & 1\\
RX\,J1757.1+6352 & 3570 & 17 57 09.7 & +63 52 38 & 17 57 09.5 & +63 52 33 & 3.31 & 0.0145 & 0.0019 & \phantom{0}\phantom{0}13.19\phantom{0}&\phantom{0} $ \phantom{0}\phantom{0}0.45\phantom{0} $ & 0.3220 & 1\\
RX\,J1757.5+6841 & 3590 & 17 57 34.1 & +68 41 22 & 17 57 34.1 & +68 41 21 & 5.46 & 0.0205 & 0.0018 & \phantom{0}\phantom{0}23.61\phantom{0}&\phantom{0} $ \phantom{0}\phantom{0}0.22\phantom{0} $ & 0.1814 & 1\\
RX\,J1757.9+6934 & 3600 & 17 57 55.2 & +69 34 23 & 17 57 55.2 & +69 34 25 & 4.35 & 0.0270 & 0.0024 & \phantom{0}\phantom{0}28.09\phantom{0}&\phantom{0} $ \phantom{0}\phantom{0}0.04\phantom{0} $ & 0.0795 & 1\\
RX\,J1757.9+6609 & \phantom{0}210 & 17 57 56.9 & +66 09 23 & 17 57 56.5 & +66 09 20 & 4.26 & 0.0025 & 0.0005 & \phantom{0}\phantom{0}\phantom{0}2.58\phantom{0}&\phantom{0} $ \phantom{0}\phantom{0}0.23\phantom{0} $ & 0.4865 & 2\\
RX\,J1758.0+6851 & 3620 & 17 58 02.4 & +68 51 46 & 17 58 03.7 & +68 51 51 & 4.87 & 0.0061 & 0.0013 & \phantom{0}\phantom{0}\phantom{0}6.69\phantom{0}&\phantom{0} $ \phantom{0}\phantom{0}0.07\phantom{0} $ & 0.1876 & 1\\
RX\,J1758.2+7020 & 3621 & 17 58 12.4 & +70 20 27 & 17 58 13.4 & +70 20 23 & 4.25 & 0.0080 & 0.0018 & \phantom{0}\phantom{0}\phantom{0}8.23\phantom{0}&\phantom{0} $ \phantom{0}\phantom{0}5.94\phantom{0} $ & 1.1400 & 1\\
RX\,J1758.2+6743 & \phantom{0}230 & 17 58 13.2 & +67 43 18 & 17 58 14.1 & +67 43 17 & 5.02 & 0.0122 & 0.0012 & \phantom{0}\phantom{0}13.55\phantom{0}&\phantom{0} $ \phantom{0}\phantom{0}0.16\phantom{0} $ & 0.2045 & 1\\
RX\,J1758.3+6906 & 3630 & 17 58 18.7 & +69 06 30 & 17 58 15.9 & +69 06 32 & 4.42 & 0.0104 & 0.0016 & \phantom{0}\phantom{0}10.91\phantom{0}&\phantom{0} $ \phantom{0}37.92\phantom{0} $ & 2.1572 & 1\\
RX\,J1758.3+6203 & \phantom{0}241 & 17 58 23.3 & +62 03 26 & 17 58 24.4 & +62 03 20 & 3.41 & 0.0078 & 0.0019 & \phantom{0}\phantom{0}\phantom{0}7.21\phantom{0}&\phantom{0} $ \phantom{0}\phantom{0}1.35\phantom{0} $ & 0.6590 & 1\\
RX\,J1758.4+6531 & \phantom{0}250 & 17 58 24.1 & +65 31 05 & 17 58 24.2 & +65 31 08 & 4.04 & 0.0195 & 0.0013 & \phantom{0}\phantom{0}19.57\phantom{0}&\phantom{0} $ \phantom{0}\phantom{0}0.68\phantom{0} $ & 0.3250 & 1\\
RX\,J1758.7+6423 & 3680 & 17 58 44.5 & +64 23 04 & 17 58 43.1 & +64 23 04 & 3.38 & 0.0054 & 0.0011 & \phantom{0}\phantom{0}\phantom{0}4.97\phantom{0}&\phantom{0} $ \phantom{0}\phantom{0}1.29\phantom{0} $ & 0.7523 & 1\\
RX\,J1758.8+6551 & \phantom{0}280 & 17 58 52.8 & +65 51 06 & 17 58 53.2 & +65 51 13 & 4.40 & 0.0030 & 0.0005 & \phantom{0}\phantom{0}\phantom{0}3.14\phantom{0}&\phantom{0} $ \phantom{0}\phantom{0}0.17\phantom{0} $ & 0.3884 & 2\\
RX\,J1758.9+6220 & 3700 & 17 58 56.5 & +62 20 31 & 17 58 56.3 & +62 20 25 & 3.33 & 0.0070 & 0.0017 & \phantom{0}\phantom{0}\phantom{0}6.38\phantom{0}&\phantom{0} $ \phantom{0}\phantom{0}1.34\phantom{0} $ & 0.6910 & 1\\
RX\,J1759.3+6335 & 3730 & 17 59 19.2 & +63 35 37 & 17 59 18.4 & +63 35 40 & 3.15 & 0.0063 & 0.0014 & \phantom{0}\phantom{0}\phantom{0}5.59\phantom{0}&\phantom{0} $ \phantom{0}\phantom{0}4.92\phantom{0} $ & 1.2354 & 1\\
RX\,J1759.7+6739 & \phantom{0}350 & 17 59 42.5 & +67 39 25 & 17 59 42.8 & +67 39 27 & 5.35 & 0.0049 & 0.0007 & \phantom{0}\phantom{0}\phantom{0}5.60\phantom{0}&\phantom{0} $ \phantom{0}\phantom{0}3.56\phantom{0} $ & 1.0830 & 1\\
RX\,J1759.7+6629 & \phantom{0}360 & 17 59 44.3 & +66 29 11 & 17 59 44.7 & +66 29 11 & 4.12 & 0.0040 & 0.0004 & \phantom{0}\phantom{0}\phantom{0}4.06\phantom{0}&\phantom{0} $ \phantom{0}\phantom{0}0.23\phantom{0} $ & 0.3990 & 1\\
RX\,J1800.1+6636 & \phantom{0}380 & 18 00 07.5 & +66 36 54 & 18 00 07.6 & +66 36 55 & 4.17 & 0.0048 & 0.0004 & \phantom{0}\phantom{0}\phantom{0}4.89\phantom{0}&\phantom{0} $ \phantom{0}\phantom{0}0.001 $ & 0.0260 & 2\\
RX\,J1800.1+6938 & \phantom{0}381 & 18 00 08.4 & +69 38 30 & 18 00 10.4 & +69 38 38 & 4.56 & 0.0055 & 0.0013 & \phantom{0}\phantom{0}\phantom{0}5.85\phantom{0}&\phantom{0} $ \phantom{0}\phantom{0}3.57\phantom{0} $ & 1.0650 & 1\\
RX\,J1800.1+6720 & \phantom{0}390 & 18 00 11.2 & +67 20 48 & 18 00 10.9 & +67 20 58 & 4.68 & 0.0036 & 0.0006 & \phantom{0}\phantom{0}\phantom{0}3.87\phantom{0}&\phantom{0} $ \phantom{0}\phantom{0}2.82\phantom{0} $ & 1.1433 & 1\\
RX\,J1800.3+6615 & \phantom{0}430 & 18 00 23.1 & +66 15 54 & 18 00 23.8 & +66 15 52 & 4.03 & 0.0048 & 0.0005 & \phantom{0}\phantom{0}\phantom{0}4.81\phantom{0}&\phantom{0} $ \phantom{0}\phantom{0}0.35\phantom{0} $ & 0.4475 & 1\\
RX\,J1800.4+7051 & 3810 & 18 00 25.2 & +70 51 55 & 18 00 25.8 & +70 51 58 & 4.25 & 0.0101 & 0.0019 & \phantom{0}\phantom{0}10.39\phantom{0}&\phantom{0} $ \phantom{0}\phantom{0}0.35\phantom{0} $ & 0.3200 & 1\\
RX\,J1800.4+6357 & 3820 & 18 00 26.2 & +63 57 19 & 18 00 26.4 & +63 57 19 & 2.72 & 0.0205 & 0.0020 & \phantom{0}\phantom{0}16.83\phantom{0}&\phantom{0} $ \phantom{0}\phantom{0}3.44\phantom{0} $ & 0.6828 & 1\\
RX\,J1800.4+6705 & \phantom{0}440 & 18 00 29.0 & +67 05 48 & 18 00 28.9 & +67 05 50 & 4.52 & 0.0041 & 0.0005 & \phantom{0}\phantom{0}\phantom{0}4.34\phantom{0}&\phantom{0} $ \phantom{0}\phantom{0}4.62\phantom{0} $ & 1.3330 & 1\\
RX\,J1801.2+6433 & 3860 & 18 01 13.2 & +64 33 22 & 18 01 14.7 & +64 33 26 & 3.43 & 0.0061 & 0.0011 & \phantom{0}\phantom{0}\phantom{0}5.65\phantom{0}&\phantom{0} $ \phantom{0}\phantom{0}2.09\phantom{0} $ & 0.8700 & 1\\
RX\,J1801.2+6902 & 3870 & 18 01 14.6 & +69 02 43 & 18 01 14.6 & +69 02 43 & 4.86 & 0.0057 & 0.0012 & \phantom{0}\phantom{0}\phantom{0}6.24\phantom{0}&\phantom{0} $ \phantom{0}\phantom{0}5.89\phantom{0} $ & 1.2700 & 1\\
RX\,J1801.2+6624 & \phantom{0}480 & 18 01 15.2 & +66 24 01 & 18 01 16.6 & +66 24 01 & 4.05 & 0.0012 & 0.0002 & \phantom{0}\phantom{0}\phantom{0}1.21\phantom{0}&\phantom{0} $ \phantom{0}\phantom{0}1.09\phantom{0} $ & 1.2500 & 1\\
RX\,J1802.0+6629 & \phantom{0}560 & 18 02 05.9 & +66 29 02 & 18 02 04.8 & +66 29 14 & 4.14 & 0.0013 & 0.0003 & \phantom{0}\phantom{0}\phantom{0}1.32\phantom{0}&\phantom{0} $ \phantom{0}\phantom{0}0.03\phantom{0} $ & 0.2650 & 1\\
RX\,J1802.1+6535 & \phantom{0}570 & 18 02 07.7 & +65 35 21 & 18 02 07.7 & +65 35 14 & 4.23 & 0.0034 & 0.0007 & \phantom{0}\phantom{0}\phantom{0}3.49\phantom{0}&\phantom{0} $ \phantom{0}\phantom{0}0.02\phantom{0} $ & 0.1513 & 1\\
RX\,J1802.3+6259 & 3910 & 18 02 19.6 & +62 59 21 & 18 02 21.5 & +62 59 14 & 3.48 & 0.0063 & 0.0014 & \phantom{0}\phantom{0}\phantom{0}5.88\phantom{0}&\phantom{0} $ \phantom{0}\phantom{0}1.39\phantom{0} $ & 0.7240 & 1\\
RX\,J1802.3+6647 & \phantom{0}590 & 18 02 22.8 & +66 47 49 & 18 02 24.5 & +66 47 35 & 4.24 & 0.0047 & 0.0005 & \phantom{0}\phantom{0}\phantom{0}4.83\phantom{0}&\phantom{0} $ \phantom{0}\phantom{0}0.19\phantom{0} $ & 0.3424 & 1\\
RX\,J1802.7+6727 & \phantom{0}630 & 18 02 47.4 & +67 27 50 & 18 02 47.8 & +67 27 41 & 4.94 & 0.0026 & 0.0006 & \phantom{0}\phantom{0}\phantom{0}2.87\phantom{0}&\phantom{0} $ \phantom{0}\phantom{0}0.02\phantom{0} $ & 0.1620 & 2\\
RX\,J1802.8+6605 & \phantom{0}640 & 18 02 51.2 & +66 05 40 & 18 02 51.3 & +66 05 42 & 4.12 & 0.0166 & 0.0009 & \phantom{0}\phantom{0}16.82\phantom{0}&\phantom{0} $ \phantom{0}\phantom{0}0.21\phantom{0} $ & 0.2070 & 1\\
RX\,J1803.4+6738 & \phantom{0}650 & 18 03 28.3 & +67 38 06 & 18 03 29.0 & +67 38 10 & 4.87 & 0.2801 & 0.0042 & \phantom{0}306.77\phantom{0}&\phantom{0} $ \phantom{0}\phantom{0}1.51\phantom{0} $ & 0.1360 & 1\\
RX\,J1803.8+6619 & \phantom{0}660 & 18 03 50.4 & +66 19 31 & 18 03 50.1 & +66 19 31 & 4.30 & 0.0064 & 0.0006 & \phantom{0}\phantom{0}\phantom{0}6.62\phantom{0}&\phantom{0} $ \phantom{0}\phantom{0}0.37\phantom{0} $ & 0.3968 & 1\\
RX\,J1804.5+6937 & 4020 & 18 04 34.2 & +69 37 33 & 18 04 34.4 & +69 37 37 & 4.76 & 0.0082 & 0.0016 & \phantom{0}\phantom{0}\phantom{0}8.90\phantom{0}&\phantom{0} $ \phantom{0}\phantom{0}1.36\phantom{0} $ & 0.6055 & 1\\
RX\,J1804.6+6846 & 4040 & 18 04 41.9 & +68 46 02 & 18 04 40.6 & +68 45 55 & 5.07 & 0.0072 & 0.0013 & \phantom{0}\phantom{0}\phantom{0}8.03\phantom{0}&\phantom{0} $ \phantom{0}\phantom{0}0.02\phantom{0} $ & 0.0969 & 2\\
RX\,J1805.2+7006 & 4080 & 18 05 16.6 & +70 06 19 & 18 05 17.8 & +70 06 22 & 4.19 & 0.0162 & 0.0024 & \phantom{0}\phantom{0}16.55\phantom{0}&\phantom{0} $ \phantom{0}\phantom{0}0.16\phantom{0} $ & 0.1874 & 1\\
RX\,J1805.4+6638 & \phantom{0}710 & 18 05 25.3 & +66 38 58 & 18 05 25.0 & +66 39 03 & 4.10 & 0.0105 & 0.0008 & \phantom{0}\phantom{0}10.61\phantom{0}&\phantom{0} $ \phantom{0}\phantom{0}0.06\phantom{0} $ & 0.1449 & 1\\
RX\,J1805.6+6624 & \phantom{0}720 & 18 05 36.1 & +66 24 52 & 18 05 36.2 & +66 24 52 & 4.10 & 0.0110 & 0.0008 & \phantom{0}\phantom{0}11.12\phantom{0}&\phantom{0} $ \phantom{0}\phantom{0}2.60\phantom{0} $ & 0.7210 & 1\\
RX\,J1805.6+6309 & 4110 & 18 05 39.0 & +63 09 36 & 18 05 40.1 & +63 09 22 & 3.00 & 0.0080 & 0.0016 & \phantom{0}\phantom{0}\phantom{0}6.92\phantom{0}&\phantom{0} $ \phantom{0}\phantom{0}0.67\phantom{0} $ & 0.5013 & 1\\
RX\,J1805.6+6432 & 4120 & 18 05 41.4 & +64 32 51 & 18 05 40.5 & +64 32 47 & 3.35 & 0.0121 & 0.0015 & \phantom{0}\phantom{0}11.07\phantom{0}&\phantom{0} $ \phantom{0}\phantom{0}2.78\phantom{0} $ & 0.7432 & 1\\
RX\,J1806.0+6940 & 4140 & 18 06 03.2 & +69 40 26 & 18 06 03.2 & +69 40 24 & 4.76 & 0.0087 & 0.0017 & \phantom{0}\phantom{0}\phantom{0}9.44\phantom{0}&\phantom{0} $ \phantom{0}\phantom{0}0.32\phantom{0} $ & 0.3214 & 1\\
RX\,J1806.2+6644 & \phantom{0}740 & 18 06 12.5 & +66 44 40 & 18 06 12.4 & +66 44 34 & 4.00 & 0.0041 & 0.0006 & \phantom{0}\phantom{0}\phantom{0}4.10\phantom{0}&\phantom{0} $ \phantom{0}\phantom{0}0.17\phantom{0} $ & 0.3482 & 1\\
RX\,J1807.7+6617 & 4270 & 18 07 47.4 & +66 17 32 & 18 07 47.4 & +66 17 31 & 3.94 & 0.0087 & 0.0008 & \phantom{0}\phantom{0}\phantom{0}8.63\phantom{0}&\phantom{0} $ \phantom{0}\phantom{0}3.82\phantom{0} $ & 0.9350 & 1\\
RX\,J1808.0+6452 & 4280 & 18 08 02.5 & +64 52 24 & 18 08 03.7 & +64 52 30 & 3.38 & 0.0211 & 0.0021 & \phantom{0}\phantom{0}19.38\phantom{0}&\phantom{0} $ \phantom{0}11.05\phantom{0} $ & 1.0360 & 1\\
RX\,J1808.8+6634 & 4390 & 18 08 49.8 & +66 34 31 & 18 08 49.6 & +66 34 29 & 3.85 & 0.0220 & 0.0013 & \phantom{0}\phantom{0}21.59\phantom{0}&\phantom{0} $ \phantom{0}\phantom{0}4.64\phantom{0} $ & 0.6970 & 1\\
RX\,J1808.8+6530 & 4400 & 18 08 51.0 & +65 30 21 & 18 08 50.8 & +65 30 19 & 3.91 & 0.0095 & 0.0013 & \phantom{0}\phantom{0}\phantom{0}9.38\phantom{0}&\phantom{0} $ \phantom{0}\phantom{0}0.26\phantom{0} $ & 0.2937 & 2\\
RX\,J1808.8+6511 & 4401 & 18 08 53.4 & +65 11 42 & 18 08 53.5 & +65 11 48 & 3.82 & 0.0043 & 0.0010 & \phantom{0}\phantom{0}\phantom{0}4.20\phantom{0}&\phantom{0} $ \phantom{0}\phantom{0}7.40\phantom{0} $ & 1.6350 & 1\\
RX\,J1809.0+6704 & 4410 & 18 09 01.0 & +67 04 21 & 18 09 00.9 & +67 04 25 & 4.09 & 0.0046 & 0.0008 & \phantom{0}\phantom{0}\phantom{0}4.65\phantom{0}&\phantom{0} $ \phantom{0}\phantom{0}0.99\phantom{0} $ & 0.6950 & 1\\
RX\,J1809.0+6800 & 4420 & 18 09 03.5 & +68 00 55 & 18 09 03.1 & +68 00 57 & 4.47 & 0.0062 & 0.0012 & \phantom{0}\phantom{0}\phantom{0}6.53\phantom{0}&\phantom{0} $ \phantom{0}\phantom{0}0.95\phantom{0} $ & 0.5946 & 1\\
RX\,J1809.0+6333 & 4430 & 18 09 05.0 & +63 33 00 & 18 09 05.5 & +63 33 09 & 2.65 & 0.0073 & 0.0017 & \phantom{0}\phantom{0}\phantom{0}5.92\phantom{0}&\phantom{0} $ \phantom{0}\phantom{0}1.04\phantom{0} $ & 0.6412 & 1\\
RX\,J1809.5+6620 & 4440 & 18 09 30.1 & +66 20 33 & 18 09 30.1 & +66 20 21 & 3.86 & 0.0061 & 0.0009 & \phantom{0}\phantom{0}\phantom{0}5.99\phantom{0}&\phantom{0} $ \phantom{0}\phantom{0}1.03\phantom{0} $ & 0.6350 & 1\\
RX\,J1809.5+6609 & 4450 & 18 09 34.8 & +66 09 06 & 18 09 34.2 & +66 09 11 & 3.85 & 0.0055 & 0.0009 & \phantom{0}\phantom{0}\phantom{0}5.40\phantom{0}&\phantom{0} $ \phantom{0}\phantom{0}2.42\phantom{0} $ & 0.9400 & 1\\
RX\,J1809.7+6837 & 4460 & 18 09 46.8 & +68 37 26 & 18 09 48.4 & +68 37 34 & 5.36 & 0.0059 & 0.0012 & \phantom{0}\phantom{0}\phantom{0}6.74\phantom{0}&\phantom{0} $ \phantom{0}\phantom{0}0.09\phantom{0} $ & 0.2173 & 1\\
RX\,J1810.0+6344 & 4490 & 18 10 04.2 & +63 44 24 & 18 10 04.4 & +63 44 26 & 2.76 & 0.0223 & 0.0026 & \phantom{0}\phantom{0}18.47\phantom{0}&\phantom{0} $ \phantom{0}\phantom{0}0.91\phantom{0} $ & 0.3770 & 1\\
RX\,J1810.3+6328 & 4501 & 18 10 23.5 & +63 28 08 & 18 10 16.9 & +63 29 14 & 2.74 & 0.0135 & 0.0025 & \phantom{0}\phantom{0}11.12\phantom{0}&\phantom{0} $ \phantom{0}\phantom{0}3.76\phantom{0} $ & 0.8380 & 1\\
RX\,J1810.4+6432 & 4520 & 18 10 24.7 & +64 32 46 & 18 10 24.2 & +64 32 54 & 3.12 & 0.0067 & 0.0015 & \phantom{0}\phantom{0}\phantom{0}5.91\phantom{0}&\phantom{0} $ \phantom{0}\phantom{0}0.17\phantom{0} $ & 0.3030 & 1\\
RX\,J1811.2+6543 & 4550 & 18 11 12.4 & +65 43 46 & 18 11 11.6 & +65 43 47 & 3.94 & 0.0127 & 0.0014 & \phantom{0}\phantom{0}12.59\phantom{0}&\phantom{0} $ \phantom{0}\phantom{0}1.15\phantom{0} $ & 0.4895 & 1\\
RX\,J1811.6+6507 & 4580 & 18 11 36.8 & +65 07 04 & 18 11 36.1 & +65 07 00 & 3.67 & 0.0197 & 0.0020 & \phantom{0}\phantom{0}18.86\phantom{0}&\phantom{0} $ \phantom{0}\phantom{0}6.54\phantom{0} $ & 0.8470 & 1\\
RX\,J1811.6+6333 & 4590 & 18 11 41.2 & +63 33 46 & 18 11 43.5 & +63 33 51 & 2.72 & 0.0084 & 0.0019 & \phantom{0}\phantom{0}\phantom{0}6.90\phantom{0}&\phantom{0} $ \phantom{0}\phantom{0}0.25\phantom{0} $ & 0.3310 & 1\\
RX\,J1812.4+6610 & 4640 & 18 12 27.0 & +66 10 46 & 18 12 26.1 & +66 10 48 & 3.87 & 0.0036 & 0.0008 & \phantom{0}\phantom{0}\phantom{0}3.54\phantom{0}&\phantom{0} $ \phantom{0}\phantom{0}0.63\phantom{0} $ & 0.6449 & 1\\
RX\,J1813.0+6644 & 4670 & 18 13 04.8 & +66 44 56 & 18 13 06.1 & +66 44 52 & 4.56 & 0.0056 & 0.0010 & \phantom{0}\phantom{0}\phantom{0}5.96\phantom{0}&\phantom{0} $ \phantom{0}\phantom{0}1.62\phantom{0} $ & 0.7680 & 1\\
RX\,J1813.1+6547 & 4680 & 18 13 09.0 & +65 47 01 & 18 13 07.7 & +65 47 04 & 3.87 & 0.0102 & 0.0013 & \phantom{0}\phantom{0}10.04\phantom{0}&\phantom{0} $ \phantom{0}\phantom{0}0.41\phantom{0} $ & 0.3489 & 1\\
RX\,J1813.1+6608 & 4690 & 18 13 10.7 & +66 08 02 & 18 13 07.9 & +66 08 09 & 3.83 & 0.0045 & 0.0009 & \phantom{0}\phantom{0}\phantom{0}4.40\phantom{0}&\phantom{0} $ \phantom{0}\phantom{0}4.75\phantom{0} $ & 1.3400 & 1\\
RX\,J1813.5+6635 & 4720 & 18 13 34.1 & +66 35 36 & 18 13 35.1 & +66 35 34 & 4.58 & 0.0038 & 0.0008 & \phantom{0}\phantom{0}\phantom{0}4.05\phantom{0}&\phantom{0} $ \phantom{0}\phantom{0}0.76\phantom{0} $ & 0.6609 & 1\\
RX\,J1813.6+6731 & 4721 & 18 13 41.5 & +67 31 50 & 18 13 43.0 & +67 32 23 & 4.67 & 0.0142 & 0.0019 & \phantom{0}\phantom{0}15.27\phantom{0}&\phantom{0} $ \phantom{0}\phantom{0}2.44\phantom{0} $ & 0.6168 & 1\\
RX\,J1813.7+6538 & 4760 & 18 13 46.6 & +65 38 21 & 18 13 45.8 & +65 38 20 & 3.91 & 0.0342 & 0.0022 & \phantom{0}\phantom{0}33.79\phantom{0}&\phantom{0} $ \phantom{0}\phantom{0}0.35\phantom{0} $ & 0.1912 & 1\\
RX\,J1813.8+6728 & 4800 & 18 13 51.0 & +67 28 10 & 18 13 50.7 & +67 28 06 & 4.63 & 0.0084 & 0.0015 & \phantom{0}\phantom{0}\phantom{0}9.00\phantom{0}&\phantom{0} $ \phantom{0}\phantom{0}0.30\phantom{0} $ & 0.3196 & 1\\
RX\,J1815.2+6658 & 4880 & 18 15 17.0 & +66 58 10 & 18 15 17.4 & +66 58 05 & 4.67 & 0.0070 & 0.0012 & \phantom{0}\phantom{0}\phantom{0}7.53\phantom{0}&\phantom{0} $ \phantom{0}\phantom{0}0.12\phantom{0} $ & 0.2287 & 2\\
RX\,J1815.3+6507 & 4890 & 18 15 19.1 & +65 07 28 & 18 15 20.0 & +65 07 14 & 3.85 & 0.0073 & 0.0015 & \phantom{0}\phantom{0}\phantom{0}7.16\phantom{0}&\phantom{0} $ \phantom{0}14.36\phantom{0} $ & 1.7234 & 1\\
RX\,J1815.4+6806 & 4910 & 18 15 24.4 & +68 06 29 & 18 15 24.9 & +68 06 32 & 4.58 & 0.0214 & 0.0022 & \phantom{0}\phantom{0}22.82\phantom{0}&\phantom{0} $ \phantom{0}\phantom{0}0.39\phantom{0} $ & 0.2390 & 1\\
RX\,J1815.8+6441 & 4930 & 18 15 52.3 & +64 41 00 & 18 15 51.7 & +64 41 03 & 3.16 & 0.0117 & 0.0020 & \phantom{0}\phantom{0}10.38\phantom{0}&\phantom{0} $ \phantom{0}\phantom{0}0.63\phantom{0} $ & 0.4116 & 1\\
RX\,J1817.5+6631 & 5020 & 18 17 32.1 & +66 31 08 & 18 17 31.3 & +66 31 11 & 4.64 & 0.0083 & 0.0012 & \phantom{0}\phantom{0}\phantom{0}8.90\phantom{0}&\phantom{0} $ \phantom{0}\phantom{0}0.67\phantom{0} $ & 0.4513 & 1\\
RX\,J1818.4+6741 & 5050 & 18 18 28.9 & +67 41 26 & 18 18 28.8 & +67 41 24 & 4.64 & 0.1125 & 0.0044 & \phantom{0}120.64\phantom{0}&\phantom{0} $ \phantom{0}\phantom{0}3.87\phantom{0} $ & 0.3140 & 1\\
RX\,J1818.7+6518 & 5080 & 18 18 46.2 & +65 18 14 & 18 18 44.8 & +65 18 10 & 4.20 & 0.0056 & 0.0013 & \phantom{0}\phantom{0}\phantom{0}5.73\phantom{0}&\phantom{0} $ \phantom{0}\phantom{0}5.82\phantom{0} $ & 1.3080 & 1\\
RX\,J1819.8+6510 & 5150 & 18 19 52.2 & +65 10 35 & 18 19 51.5 & +65 10 37 & 4.21 & 0.0173 & 0.0020 & \phantom{0}\phantom{0}17.72\phantom{0}&\phantom{0} $ \phantom{0}\phantom{0}0.18\phantom{0} $ & 0.1894 & 1\\
RX\,J1819.9+6628 & 5190 & 18 19 59.9 & +66 28 25 & 18 19 59.0 & +66 28 30 & 4.79 & 0.0049 & 0.0012 & \phantom{0}\phantom{0}\phantom{0}5.33\phantom{0}&\phantom{0} $ \phantom{0}\phantom{0}5.07\phantom{0} $ & 1.2740 & 1\\
RX\,J1820.5+6620 & 5230 & 18 20 32.9 & +66 20 29 & 18 20 32.9 & +66 20 20 & 4.71 & 0.0075 & 0.0015 & \phantom{0}\phantom{0}\phantom{0}8.09\phantom{0}&\phantom{0} $ \phantom{0}\phantom{0}0.80\phantom{0} $ & 0.5057 & 1\\
RX\,J1820.5+6930 & 5240 & 18 20 35.3 & +69 30 04 & 18 20 33.4 & +69 30 15 & 6.46 & 0.0160 & 0.0025 & \phantom{0}\phantom{0}19.77\phantom{0}&\phantom{0} $ \phantom{0}\phantom{0}0.24\phantom{0} $ & 0.2051 & 1\\
RX\,J1821.6+6543 & 5300 & 18 21 38.8 & +65 43 04 & 18 21 40.0 & +65 43 10 & 4.26 & 0.0171 & 0.0020 & \phantom{0}\phantom{0}17.63\phantom{0}&\phantom{0} $ \phantom{0}\phantom{0}0.39\phantom{0} $ & 0.2666 & 2\\
RX\,J1821.6+6328 & 5310 & 18 21 39.6 & +63 28 27 & 18 21 38.8 & +63 28 26 & 3.10 & 0.0282 & 0.0035 & \phantom{0}\phantom{0}24.80\phantom{0}&\phantom{0} $ \phantom{0}\phantom{0}1.13\phantom{0} $ & 0.3656 & 1\\
RX\,J1821.9+6654 & 5321 & 18 21 55.7 & +66 54 34 & 18 21 56.5 & +66 54 26 & 5.07 & 0.0056 & 0.0011 & \phantom{0}\phantom{0}\phantom{0}6.24\phantom{0}&\phantom{0} $ \phantom{0}\phantom{0}0.01\phantom{0} $ & 0.0873 & 1\\
RX\,J1821.9+6420 & 5340 & 18 21 57.4 & +64 20 51 & 18 21 57.1 & +64 20 37 & 3.77 & 1.0710 & 0.0140 & 1039.63\phantom{0}&\phantom{0} $ \phantom{0}29.32\phantom{0} $ & 0.2970 & 1\\
RX\,J1821.9+6818 & 5330 & 18 21 58.8 & +68 18 42 & 18 21 59.4 & +68 18 42 & 4.85 & 0.0073 & 0.0017 & \phantom{0}\phantom{0}\phantom{0}7.98\phantom{0}&\phantom{0} $ \phantom{0}15.30\phantom{0} $ & 1.6920 & 1\\
RX\,J1823.3+6419 & 5400 & 18 23 20.0 & +64 19 23 & 18 23 19.2 & +64 19 32 & 3.32 & 0.0219 & 0.0026 & \phantom{0}\phantom{0}19.95\phantom{0}&\phantom{0} $ \phantom{0}\phantom{0}2.70\phantom{0} $ & 0.5766 & 1\\
RX\,J1823.6+6847 & 5411 & 18 23 38.6 & +68 47 40 & 18 23 39.4 & +68 47 46 & 5.85 & 0.0085 & 0.0019 & \phantom{0}\phantom{0}10.08\phantom{0}&\phantom{0} $ \phantom{0}\phantom{0}0.13\phantom{0} $ & 0.2071 & 1\\
RX\,J1823.9+6719 & 5440 & 18 23 54.6 & +67 19 41 & 18 23 54.7 & +67 19 36 & 4.75 & 0.0079 & 0.0015 & \phantom{0}\phantom{0}\phantom{0}8.56\phantom{0}&\phantom{0} $ \phantom{0}\phantom{0}0.65\phantom{0} $ & 0.4536 & 1\\
\phantom{$^\star$}RX\,J1824.7+6509$^\star$ & 5500 & 18 24 46.9 & +65 09 24 & 18 24 47.3 & +65 09 25 & 3.97 & 0.1269 & 0.0048 & \phantom{0}126.39\phantom{0}&\phantom{0} $ \phantom{0}\phantom{0}3.73\phantom{0} $ & 0.3030 & 1\\
RX\,J1825.7+6905 & 5530 & 18 25 46.3 & +69 05 51 & 18 25 47.3 & +69 05 54 & 6.66 & 0.0362 & 0.0036 & \phantom{0}\phantom{0}45.27\phantom{0}&\phantom{0} $ \phantom{0}\phantom{0}0.09\phantom{0} $ & 0.0888 & 1\\
RX\,J1826.6+6706 & 5550 & 18 26 38.3 & +67 06 47 & 18 26 37.5 & +67 06 44 & 5.48 & 0.0212 & 0.0023 & \phantom{0}\phantom{0}24.46\phantom{0}&\phantom{0} $ \phantom{0}\phantom{0}0.64\phantom{0} $ & 0.2870 & 1\\
RX\,J1827.2+6549 & 5560 & 18 27 15.3 & +65 49 21 & 18 27 13.9 & +65 49 20 & 5.30 & 0.0071 & 0.0016 & \phantom{0}\phantom{0}\phantom{0}8.07\phantom{0}&\phantom{0} $ \phantom{0}\phantom{0}6.97\phantom{0} $ & 1.2250 & 1\\
RX\,J1827.5+6431 & 5590 & 18 27 33.6 & +64 31 38 & 18 27 33.8 & +64 31 44 & 4.16 & 0.0141 & 0.0022 & \phantom{0}\phantom{0}14.35\phantom{0}&\phantom{0} $ \phantom{0}\phantom{0}0.03\phantom{0} $ & 0.0977 & 1\\
RX\,J1828.1+6709 & 5601 & 18 28 06.6 & +67 09 23 & 18 28 06.7 & +67 09 17 & 6.02 & 0.0085 & 0.0016 & \phantom{0}\phantom{0}10.20\phantom{0}&\phantom{0} $ \phantom{0}\phantom{0}4.61\phantom{0} $ & 0.9430 & 1\\
RX\,J1828.2+6403 & 5620 & 18 28 13.7 & +64 03 31 & 18 28 14.2 & +64 03 28 & 3.83 & 0.0119 & 0.0023 & \phantom{0}\phantom{0}11.64\phantom{0}&\phantom{0} $ \phantom{0}\phantom{0}0.03\phantom{0} $ & 0.0963 & 1\\
RX\,J1828.7+6953 & 5670 & 18 28 47.9 & +69 53 58 & 18 28 49.3 & +69 54 00 & 7.02 & 0.0271 & 0.0033 & \phantom{0}\phantom{0}34.60\phantom{0}&\phantom{0} $ \phantom{0}\phantom{0}0.11\phantom{0} $ & 0.1100 & 1\\
RX\,J1828.8+6452 & 5680 & 18 28 48.6 & +64 52 50 & 18 28 48.3 & +64 53 00 & 4.42 & 0.0134 & 0.0020 & \phantom{0}\phantom{0}14.05\phantom{0}&\phantom{0} $ \phantom{0}\phantom{0}5.25\phantom{0} $ & 0.8730 & 1\\
RX\,J1829.0+6433 & 5690 & 18 29 00.6 & +64 33 49 & 18 29 00.4 & +64 33 51 & 4.30 & 0.0183 & 0.0023 & \phantom{0}\phantom{0}18.94\phantom{0}&\phantom{0} $ \phantom{0}\phantom{0}1.00\phantom{0} $ & 0.3880 & 1\\
RX\,J1829.5+6631 & 5740 & 18 29 35.4 & +66 31 19 & 18 29 35.4 & +66 31 23 & 6.21 & 0.0066 & 0.0016 & \phantom{0}\phantom{0}\phantom{0}8.03\phantom{0}&\phantom{0} $ \phantom{0}\phantom{0}1.68\phantom{0} $ & 0.6898 & 1\\
RX\,J1829.7+6749 & 5750 & 18 29 43.4 & +67 49 09 & 18 29 42.2 & +67 49 12 & 6.19 & 0.0158 & 0.0023 & \phantom{0}\phantom{0}19.18\phantom{0}&\phantom{0} $ \phantom{0}\phantom{0}1.66\phantom{0} $ & 0.4783 & 1\\
RX\,J1830.0+6645 & 5790 & 18 30 01.4 & +66 45 23 & 18 30 02.0 & +66 45 23 & 6.38 & 0.0142 & 0.0021 & \phantom{0}\phantom{0}17.45\phantom{0}&\phantom{0} $ \phantom{0}\phantom{0}0.46\phantom{0} $ & 0.2889 & 1\\
RX\,J1830.1+6425 & 5800 & 18 30 07.5 & +64 25 28 & 18 30 06.1 & +64 25 29 & 3.90 & 0.0104 & 0.0019 & \phantom{0}\phantom{0}10.26\phantom{0}&\phantom{0} $ \phantom{0}\phantom{0}1.69\phantom{0} $ & 0.6253 & 1\\
RX\,J1832.0+6542 & 5890 & 18 32 01.5 & +65 42 35 & 18 32 01.3 & +65 42 33 & 5.41 & 0.0239 & 0.0026 & \phantom{0}\phantom{0}27.42\phantom{0}&\phantom{0} $ \phantom{0}\phantom{0}0.74\phantom{0} $ & 0.2908 & 1\\
RX\,J1832.0+6447 & 5900 & 18 32 04.2 & +64 47 01 & 18 32 01.2 & +64 47 08 & 4.33 & 0.0177 & 0.0025 & \phantom{0}\phantom{0}18.37\phantom{0}&\phantom{0} $ \phantom{0}12.65\phantom{0} $ & 1.1180 & 1\\
RX\,J1832.4+6402 & 5930 & 18 32 25.2 & +64 02 02 & 18 32 24.1 & +64 02 10 & 4.93 & 0.0106 & 0.0022 & \phantom{0}\phantom{0}11.68\phantom{0}&\phantom{0} $ \phantom{0}\phantom{0}0.59\phantom{0} $ & 0.3826 & 1\\
RX\,J1832.4+6438 & 5940 & 18 32 25.2 & +64 38 15 & 18 32 24.2 & +64 38 23 & 4.27 & 0.0140 & 0.0021 & \phantom{0}\phantom{0}14.45\phantom{0}&\phantom{0} $ \phantom{0}\phantom{0}1.62\phantom{0} $ & 0.5335 & 1\\
RX\,J1833.0+6344 & 5990 & 18 33 02.8 & +63 44 17 & 18 33 01.3 & +63 44 35 & 4.93 & 0.0139 & 0.0024 & \phantom{0}\phantom{0}15.31\phantom{0}&\phantom{0} $ \phantom{0}\phantom{0}1.88\phantom{0} $ & 0.5535 & 1\\
RX\,J1835.0+6526 & 6070 & 18 35 04.8 & +65 26 44 & 18 35 06.0 & +65 26 47 & 5.52 & 0.0206 & 0.0026 & \phantom{0}\phantom{0}23.85\phantom{0}&\phantom{0} $ \phantom{0}\phantom{0}1.41\phantom{0} $ & 0.4083 & 1\\
RX\,J1835.1+6342 & 6080 & 18 35 08.0 & +63 42 33 & 18 35 10.0 & +63 43 14 & 4.93 & 0.0228 & 0.0033 & \phantom{0}\phantom{0}25.12\phantom{0}&\phantom{0} $ \phantom{0}11.40\phantom{0} $ & 0.9445 & 1\\
RX\,J1835.1+6733 & 6090 & 18 35 10.3 & +67 33 54 & 18 35 09.0 & +67 33 58 & 6.59 & 0.0075 & 0.0017 & \phantom{0}\phantom{0}\phantom{0}9.34\phantom{0}&\phantom{0} $ \phantom{0}\phantom{0}2.37\phantom{0} $ & 0.7460 & 1\\
RX\,J1836.4+6602 & 6180 & 18 36 28.2 & +66 02 40 & 18 36 28.7 & +66 02 37 & 7.08 & 0.0130 & 0.0024 & \phantom{0}\phantom{0}16.65\phantom{0}&\phantom{0} $ \phantom{0}\phantom{0}0.16\phantom{0} $ & 0.1858 & 1\\
RX\,J1836.6+6719 & 6200 & 18 36 36.1 & +67 19 04 & 18 36 36.8 & +67 19 12 & 8.03 & 0.0084 & 0.0018 & \phantom{0}\phantom{0}11.29\phantom{0}&\phantom{0} $ \phantom{0}\phantom{0}0.25\phantom{0} $ & 0.2693 & 1\\
RX\,J1838.1+6649 & 6280 & 18 38 09.1 & +66 49 26 & 18 38 10.0 & +66 49 22 & 7.17 & 0.0083 & 0.0019 & \phantom{0}\phantom{0}10.69\phantom{0}&\phantom{0} $ \phantom{0}\phantom{0}0.28\phantom{0} $ & 0.2879 & 1\\
RX\,J1838.8+6432 & 6300 & 18 38 51.9 & +64 32 21 & 18 38 53.1 & +64 32 23 & 5.43 & 0.0081 & 0.0019 & \phantom{0}\phantom{0}\phantom{0}9.31\phantom{0}&\phantom{0} $ \phantom{0}\phantom{0}0.77\phantom{0} $ & 0.4700 & 1\\
RX\,J1839.2+6711 & 6301 & 18 39 16.9 & +67 11 12 & 18 39 16.5 & +67 11 06 & 8.24 & 0.0063 & 0.0015 & \phantom{0}\phantom{0}\phantom{0}8.55\phantom{0}&\phantom{0} $ \phantom{0}\phantom{0}1.94\phantom{0} $ & 0.7130 & 1\\
RX\,J1839.3+6544 & 6340 & 18 39 18.5 & +65 44 42 & 18 39 18.3 & +65 44 35 & 5.77 & 0.0095 & 0.0023 & \phantom{0}\phantom{0}11.20\phantom{0}&\phantom{0} $ \phantom{0}\phantom{0}0.02\phantom{0} $ & 0.0820 & 1\\
RX\,J1841.3+6321 & 6450 & 18 41 18.9 & +63 21 36 & 18 41 20.0 & +63 21 42 & 5.32 & 0.0096 & 0.0022 & \phantom{0}\phantom{0}10.94\phantom{0}&\phantom{0} $ \phantom{0}15.55\phantom{0} $ & 1.4990 & 1\\
RX\,J1842.2+6204 & 6452 & 18 42 14.8 & +62 04 24 & 18 42 15.6 & +62 04 24 & 5.25 & 0.0146 & 0.0032 & \phantom{0}\phantom{0}16.54\phantom{0}&\phantom{0} $ \phantom{0}\phantom{0}0.56\phantom{0} $ & 0.3203 & 1\\
RX\,J1842.5+6809 & 6490 & 18 42 33.0 & +68 09 30 & 18 42 33.3 & +68 09 25 & 6.35 & 0.0511 & 0.0044 & \phantom{0}\phantom{0}62.70\phantom{0}&\phantom{0} $ \phantom{0}\phantom{0}5.33\phantom{0} $ & 0.4750 & 1\\
RX\,J1842.9+6241 & 6491 & 18 42 56.4 & +62 41 44 & 18 42 55.2 & +62 41 49 & 5.32 & 0.0502 & 0.0048 & \phantom{0}\phantom{0}57.21\phantom{0}&\phantom{0} $ \phantom{0}\phantom{0}0.10\phantom{0} $ & 0.0835 & 1\\
RX\,J1843.3+6653 & 6520 & 18 43 22.5 & +66 53 21 & 18 43 20.9 & +66 53 29 & 7.33 & 0.0110 & 0.0022 & \phantom{0}\phantom{0}14.28\phantom{0}&\phantom{0} $ \phantom{0}\phantom{0}0.50\phantom{0} $ & 0.3273 & 1\\
RX\,J1843.9+6821 & 6540 & 18 43 55.7 & +68 21 11 & 18 43 54.1 & +68 21 01 & 6.08 & 0.0200 & 0.0032 & \phantom{0}\phantom{0}24.10\phantom{0}&\phantom{0} $ \phantom{0}\phantom{0}1.12\phantom{0} $ & 0.3688 & 1\\
RX\,J1844.3+6431 & 6542 & 18 44 23.4 & +64 31 31 & 18 44 21.8 & +64 31 46 & 6.00 & 0.0105 & 0.0022 & \phantom{0}\phantom{0}12.59\phantom{0}&\phantom{0} $ \phantom{0}\phantom{0}1.13\phantom{0} $ & 0.4870 & 1\\
RX\,J1844.4+6236 & 6543 & 18 44 26.9 & +62 36 12 & 18 44 26.4 & +62 36 14 & 5.32 & 0.0274 & 0.0039 & \phantom{0}\phantom{0}31.22\phantom{0}&\phantom{0} $ \phantom{0}\phantom{0}1.03\phantom{0} $ & 0.3172 & 1\\
RX\,J1844.4+6248 & 6544 & 18 44 27.5 & +62 48 27 & 18 44 26.2 & +62 48 29 & 5.32 & 0.0239 & 0.0034 & \phantom{0}\phantom{0}27.24\phantom{0}&\phantom{0} $ \phantom{0}67.62\phantom{0} $ & 1.8800 & 1\\
RX\,J1844.9+6813 & 6570 & 18 44 54.0 & +68 13 23 & 18 44 54.1 & +68 13 17 & 5.84 & 0.0250 & 0.0036 & \phantom{0}\phantom{0}29.64\phantom{0}&\phantom{0} $ \phantom{0}\phantom{0}0.92\phantom{0} $ & 0.3097 & 1\\